\documentclass{aastex}
\usepackage[onecolumn,numberedappendix]{emulateapj5}

\usepackage{epsfig}
\usepackage{rotating}
\tightenlines

\newcommand     \kms    {\,{\rm km~s}^{-1}}

\newcommand     \s      {\,{\rm s}}

\newcommand{\smyr}      {{ M_\odot\ \rm yr^{-1}}}
\newcommand{\sm}        {{ M_\odot}}

\newcommand{\beq}       {\begin{equation}}
\newcommand{\eeq}       {\end{equation}}
\newcommand{\beqa}      {\begin{eqnarray}}
\newcommand{\eeqa}      {\end{eqnarray}}
\newcommand{\e}         {$^{-1}$}
\newcommand{\ee}        {$^{-2}$}
\newcommand{\eee}       {$^{-3}$}

\newcommand{\fkep}      {f_{\rm Kep}}
\newcommand{\krho}      {{k_\rho}}

\newcommand{\mds}       {\dot m_*}
\newcommand{\msd}       {m_{*d}}
\newcommand{\msdo}       {m_{*d,\, 0}}
\newcommand{\mdsd}      {\dot m_{*d}}
\newcommand{\mdsdo}      {\dot m_{*d,\, 0}}
\newcommand{\tff}       {t_{\rm ff}}
\newcommand{\mcore}     {M_{\rm core}}

\newcommand{\htwo}      {H$_2$}

\newcommand{\muh}       {\mu_{\rm H}}
\newcommand{\scr}       {S_{\rm cr}}

\newcommand{\atwo}      {\alpha^{(2)}}

\newcommand{\lal}       {Lyman-$\alpha$}
\newcommand{\btl}       {\bar\tau_L}
\newcommand{\calf}      {{\cal F}}
\newcommand{\crit}      {{\rm crit}}
\newcommand{\dnd}       {\Delta\nu_D}
\newcommand{\dvd}       {\Delta v_D}
\newcommand{\dvds}      {\Delta v_{D,\, 6}}
\newcommand{\iso}       {{\rm iso}}
\newcommand{\necr}      {n_{e,\,\rm cr}}
\newcommand{\rhk}       {\rho_{\rm HK}}
\newcommand{\taueff}    {\bar\tau_{\rm eff}}
\newcommand{\vecfrad}   {{\bf f}_{\rm rad}}
\newcommand{\prad}      {{P}_{\rm rad}}
\newcommand{\pradi}     {{P}_{\rm rad,\, iso}}
\newcommand{\pradrr}    {{P}_{{\rm rad},rr}}
\newcommand{\urad}      {{u}_{\rm rad}}
\newcommand{\uradi}      {{u}_{\rm rad,\, iso}}
\newcommand{\rvecprad}  {{\bf P}_{\rm rad}}
\newcommand{\vecF}      {{\bf F}}
\newcommand{\vecFhat}   {{\bf \hat F}}
\newcommand{\vecnabla}  {{\bf\nabla}}
\newcommand{\vecnhat}	{{\bf \hat n}}

\newcommand{\avg}[1]    {\langle #1 \rangle}
\newcommand{\pbyp}[2]   {\frac{\partial #1}{\partial #2}}
\newcommand{\cgc}       {c_{gc}}
\newcommand{\esd}       {\epsilon_{*d}}
\newcommand{\esdb}       {\bar\epsilon_{*d}}
\newcommand{\fsh}       {f_{\rm sh}}
\newcommand{\mdst}      {\dot m_{*,\,-3}}
\newcommand{\mdsdt}     {\dot m_{*d,\,-3}}
\newcommand{\mst}       {m_{*,\,2}}
\newcommand{\msdt}      {m_{*d,\,2}}

\newcommand{\ssc}       {\left(\frac{\Sigma}{\Sigma_c}\right)}
\newcommand{\kapr}      {\kappa_{\rm R}}
\newcommand{\kht}       {\kappa_{\rm T}}
\newcommand{\krs}       {\kappa_{{\rm R}, s}}
\newcommand{\ledd}      {L_{\rm Edd}}
\newcommand{\ncr}       {n_{\rm cr}}
\newcommand{\phiedd} {\phi_{\rm Edd}}
\newcommand{\rcr}       {r_{\rm cr}}
\newcommand{\rtwo}	{r_{\rm HII}}

\newcommand{\teff}      {T_{\rm eff}}

\newcommand{\tyr}       {t_{\rm yr}}
\newcommand{\vff}       {v_{\rm ff}}

\newcommand{\zph}       {z_{\rm ph}}
\newcommand{\zsr}       {z_{sr}}

\newlength{\figwidth}
\countdef\decade=200
\advance\decade by \year
\countdef\hours=201
\advance\hours by \time
\divide\hours by 60
\countdef\mins=202
\advance\mins by \hours
\multiply\mins by 60
\multiply\hours by 100
\countdef\miltime=203
\advance\miltime by \hours
\advance\miltime by \time
\advance\miltime by -\mins
\newcommand     \todayd{\number\decade.\number\month.\number\day.\number\miltime}
\begin{document}

\title{The Formation of the First Stars II.\\ 
Radiative Feedback Processes and Implications for the Initial Mass Function
        \\
        {\small DRAFT: \today}}

\author{Christopher F. McKee$^1$ and Jonathan C. Tan$^2$ 
        }

\affil{1. Departments of Physics \& Astronomy, University of California,
Berkeley, CA 94720, USA.\\cmckee@astro.berkeley.edu}
\affil{2. Department of Astronomy, University of Florida, Gainesville, FL 32611, USA.\\jt@astro.ufl.edu}

\begin{abstract}

  We consider the radiative feedback processes that operate during the
  formation of the first stars, including the photodissociation of
  H$_2$, \lal\ radiation pressure, formation and expansion of an
  \ion{H}{2} region, and disk photoevaporation. These processes may
  inhibit continued accretion once the stellar mass has reached a
  critical value, and we evaluate this mass separately for each
  process. Photodissociation of H$_2$ in the local dark matter
  minihalo occurs relatively early in the growth of the protostar, but
  we argue this does not affect subsequent accretion since by this
  time the depth of the potential is large enough for accretion to be
  mediated by atomic cooling. However, neighboring starless minihalos
  can be affected. Ionization creates an \ion{H}{2} region in the
  infalling envelope above and below the accretion disk. \lal\
  radiation pressure acting at the boundary of the \ion{H}{2} region
  is effective at reversing infall from narrow polar directions when
  the star reaches $\sim 20-30 M_\odot$, but cannot prevent infall
  from other directions. Expansion of the \ion{H}{2} region beyond the
  gravitational escape radius for ionized gas occurs at masses $\sim
  50-100\sm$, depending on the accretion rate and angular momentum of
  the inflow. However, again, accretion from the equatorial regions
  can continue since the neutral accretion disk has a finite thickness
  and shields a substantial fraction of the accretion envelope from
  direct ionizing flux. At higher stellar masses, $\sim 140\sm$ in the
  fiducial case, the combination of declining accretion rates and
  increasing photoevaporation-driven mass loss from the disk act to
  effectively halt the increase in the protostellar mass. We identify
  this process as the mechanism that terminates the growth of
  Population III stars (i.e., stars with primordial composition) that
  have not been affected by prior star formation (Population III.1
  stars).  We discuss the implications of our results for the initial
  mass function of these stars. In the Appendix we develop approximate
  solutions to a number of problems relevant to the formation of the
  first stars: the effect of Rayleigh scattering on line profiles in
  media of very large optical depth; the intensity of Lyman-$\alpha$
  radiation in very opaque media; an approximate determination of the
  radiative acceleration in terms of the gradient of a modified
  radiation pressure; the determination of the flux of radiation in a
  shell with an arbitrary distribution of opacity; and the vertical
  structure of an accretion disk supported by gas pressure with
  constant opacity.
\end{abstract}


\section{Introduction
        \label{S:intro}}

        There has been substantial recent progress in our theoretical
understanding of how the first stars formed (Bromm \& Larson 2004).
In marked contrast to the case for contemporary star formation, the
initial conditions for the formation of the first stars are believed
to be relatively well understood: they are determined by the growth of
small-scale gravitational instabilities from cosmological fluctuations
in a cold dark matter universe. The first stars are expected to form
at redshifts $z\sim 10-50$ in dark matter halos of mass $\sim 10^6
M_\odot$ (Tegmark et al. 1997).  In the absence of any elements heavier
than helium (other than trace amounts of lithium) the chemistry and
thermodynamics of the gas are very simple (Abel, Bryan, \& Norman
2002, hereafter ABN; Bromm, Coppi, \& Larson 2002, hereafter BCL02).
There are no dust grains to couple the gas to radiation emitted by the
protostar.  There are no previous generations of stars to roil the gas
out of which the stars form, nor is there any radiation other than the
cosmic background radiation.  Existing calculations have assumed that
there were no significant primordial magnetic fields, thereby
eliminating a major complication that occurs in contemporary star
formation.  However, even in the absence of significant primordial
fields, it is possible that magnetic fields could have been generated
in the accretion disk surrounding a primordial protostar (Tan \&
Blackman 2004), although even in this case the magnetic fields become
dynamically important only after the star formation process is well
underway, and they do not affect the initial conditions.  Given this
relative simplicity, there is some confidence in the results of
numerical simulations that have followed the collapse of cosmological
scale perturbations down to almost stellar dimensions 
(ABN; Bromm, Coppi, \& Larson 1999; Yoshida et al. 2006; O'Shea \&
Norman 2007). 
This confidence is strengthened by the fact that it
appears to be the microphysics of $\rm H_2$ cooling that determines
the types of baryonic structures that are formed, and not, for
example, the details of the initial power spectrum of fluctuations in
dark matter density. The results of these simulations suggest that the
initial gas cores out of which stars form are quite massive, $\mcore
\sim 100 - 1000\sm$.

        The observational imprint of the first stars depends on their
mass: These stars were likely to be of critical importance in
reionizing the universe, in producing the first metals, and in
creating the first stellar-mass black holes.  The number of ionizing
photons emitted per baryon depends on the stellar mass for $m_*\la 300
M_\odot$ (Bromm, Kudritzki, \& Loeb 2001). The hardness of the
radiation field is also sensitive to the mass (e.g. Tumlinson \& Shull
2000; Schaerer 2002), so that He reionization can be affected. The
effectiveness of the first stars in enriching the intergalactic medium
(IGM) with metals and in producing the first stellar-mass black holes
also depends sensitively on the mass of the star. A potential
simplification in assessing these effects is that massive primordial
stars are thought to have much smaller mass-loss rates than
contemporary massive stars (Kudritzki 2002), so that the mass at core
collapse should be quite similar to the initial mass.  However,
Meynet, Ekstr\"om, \& Maeder (2006) have argued that if rotation is
allowed for, then mass loss can be significant.  Heger \& Woosley
(2002) showed that stars exploding as supernovae above about $260
M_\odot$ and between 40 and $140 M_\odot$ should collapse directly to
black holes, and they argued that such stars would providing
relatively little metal enrichment.  However, Ohkubo et al. (2007)
followed the collapse and explosion of $500 M_\odot$ and $1000
M_\odot$ stars in two dimensions and concluded that substantial amount
of metals could be ejected; they proposed that such supernovae could
produce intermediate-mass black holes.  Heger \& Woosley (2002) also
showed that for $140\la m_*\la 260 M_\odot$, the pair instability
leads to explosive O and Si burning that completely disrupts the star,
leaving no remnant and ejecting large quantities of heavy
elements. Such supernovae produce a dramatic odd-even effect in the
nuclei produced.  Stars below $\sim 40 \sm$ are expected to form
neutron stars, with more normal enrichment rates. In principle,
metallicity determinations from high-redshift absorption line systems
(Schaye et al. 2003; Norman, O'Shea, \& Paschos 2004) or from very
metal poor local stars (Beers \& Christlieb 2005) can constrain the
initial mass function (IMF) of the early generations of stars.
Indeed, based on observations such as these, Daigne et al. (2004)
argue that the stars responsible for reionizing the universe mostly
likely had masses $\la 100\sm$, and Tumlinson (2006) concludes that
stars above $140 M_\odot$ could have contributed at most 10\% of the
iron observed in extremely metal poor stars (those with
[Fe/H]~$<10^{-3}$---Beers \& Christlieb 2005).

	In discussing the formation of the first stars, the terms
``first stars'' and ``Population III stars'' are often used
interchangeably, but this can lead to confusion. To be precise, we
shall follow the conventions suggested by one of us at the First Stars
III conference (O'Shea et al. 2008) and define Population III stars as
those stars with a metallicity sufficiently low that it has no effect
on either the formation or the evolution of the stars. The value of
the critical metallicity for star formation---i.e., the value below
which the metals do not influence star formation---is uncertain, with
estimates ranging from $\sim 10^{-6}Z_\odot$ if the cooling is
dominated by small dust grains that contain a significant fraction of
the metals (Omukai et al. 2005) to $\sim 10^{-3.5}Z_\odot$ if the
cooling is dominated by the fine structure lines of C and O and there
is negligible \htwo\ (Bromm \& Loeb 2003); 
if \htwo\ cooling is included, Jappsen et al. (2007) argue that there is
no critical metallicity for gas-phase metals.  It is possible that
values of the metallicity even less than $10^{-6}Z_\odot$ could
significantly affect the evolution of primordial stars (Meynet,
private communication). Among Population III stars, we distinguish
between the first and second generations, termed Population III.1 and
III.2, respectively: The initial conditions for the formation of
Population III.1 stars are determined solely by cosmological
fluctuations whereas those for Population III.2 stars are
significantly affected by other stars.  It is likely that Population
III.1 stars have a primordial composition, since it is hard to see how
the gas out of which they form could be contaminated by even small
trace amounts of metals without having been affected by radiation from
the star that produced the metals.  Stars affect the initial
conditions for the formation of Population III.2 stars primarily by
radiation, both ionizing radiation and Lyman-Werner band radiation
that destroys \htwo\ molecules. The latter reduces the cooling
efficiency of the gas, allowing compression to heat up to the point
that it begins to become collisionally ionized; shocks associated with
\ion{H}{2} regions and supernova remnants can also ionize the gas. Once the
gas has been partially ionized, HD cooling can become important,
reducing the characteristic star-formation mass (Uehara \& Inutsuka
2000). Greif \& Bromm (2006) use the term ``Population II.5" to
describe stars that form from gas in which HD cooling is important,
whereas in our terminology these would be Population III.2 stars, but
we believe that it is better to describe all essentially metal-free
stars as ``Population III." It should be noted that our definition of
Population III.2 stars includes all Population III stars that were
significantly affected by previous generations of star formation, even
if that did not result in significant HD production.  Those Population
III.2 stars that form out of gas that is enriched in HD will typically
be less massive than Population III.1 stars---by about a factor 10
according to Greif \& Bromm (2006). They infer that Population III.1
stars are relatively rare, constituting about 10\% by mass of all
Population III stars.

	In this paper we wish to estimate the characteristic mass of
the first generation of stars (Population III.1). Even if they are
relatively rare, they are critical in setting the initial conditions
for the star formation that followed, and therefore in determining the
reionization of the universe and the production of the first metals
and the first stellar mass black holes. For contemporary star
formation, it is believed that the IMF is set by a combination of
gravitational fragmentation in a turbulent medium (Elmegreen 1997;
Padoan \& Nordlund 2002) and feedback effects.  The characteristic
stellar mass is of order the Bonnor-Ebert mass, $m_{\rm BE}\propto
T^{3/2}/\rho^{1/2}$.  However, not all of the initial core mass is
incorporated into the final star, since contemporary protostars have
powerful outflows that eject some of the core mass (Nakano, Hasegawa,
\& Norman 1995; Matzner \& McKee 2000). There are a number of feedback
effects that occur in contemporary massive star formation (Larson \&
Starrfield 1971), particularly radiation pressure on dust and
photoionization associated with the growth of an \ion{H}{2} region.
It remains unclear whether the upper limit on the contemporary IMF is
set by feedback or by instabilities that afflict very massive stars.

        The clumps out of which the first stars form have total
masses, including dark matter, of order $10^6 M_\odot$; these objects
are typically referred to as minihalos.  Cooling by trace amounts of
\htwo\ generally leads to the formation of a gravitationally unstable
core of baryonic mass $\sim 10^{2-3} \sm$ (ABN). In contrast to
contemporary star-forming regions, the turbulence in this gas is
subsonic, due to the higher temperatures and the lack of internal and
external sources of turbulence.  As a result, gravitational
fragmentation is much less effective---indeed, numerical simulations
show no evidence for it (e.g., ABN, Yoshida et al. 2006,
but see Clark, Glover, \& Klessen 2008),
and analytic calculations 
(Ripamonti \& Abel 2004), including those that consider disk
fragmentation (Tan \& Blackman 2004), show no evidence for
fragmentation either. 
It therefore appears that the mass of the first stars is likely to be
set by feedback effects.

        Feedback effects can be classified as either radiative or
kinetic. Kinetic feedback includes protostellar outflows and
main-sequence winds. In contemporary star formation, outflows
are believed to be hydromagnetically driven; in this paper,
we assume that the magnetic fields associated with the protostar
are too weak or too tangled to drive a strong outflow 
(see Tan \& Blackman 2004 for a more extensive discussion of the
effect of these outflows).
Due to the absence of metals, main-sequence winds are very weak in the
absence of rotation (Kudritzki 2002), although they could become
important in the later stages of evolution if the star is rapidly
rotating. Since we are primarily interested in the early stages of
evolution of the star, we neglect kinetic feedback (i.e., outflows and
winds).

        Radiative feedback includes radiation pressure,
photoionization heating, and photodissociation. Radiation pressure can
be due to continuum radiation or to resonance line scattering; the
continuum radiation pressure can be due to electron scattering or to
photoionization. Photoionization also leads to heating, which unbinds
gas beyond the gravitational radius $r_g\equiv Gm_*/c_s^2$, where
$c_s$ is the isothermal sound speed of the gas. If the gas is
initially in a disk, this process is termed photoevaporation. Finally,
photodissociation destroys \htwo\, the dominant coolant in neutral
primordial gas.

        Most previous work has focused on the effects of continuum
radiation pressure and photoionization heating in limiting the mass of
primordial stars.  Omukai \& Palla (2001; 2003) focused on electron
scattering, which leads to the Eddington limit on an accreting mass.
For the case of spherical accretion at a rate of $4.4\times
10^{-3}\smyr$, radiation pressure first becomes important at around
$80\sm$, leading to a dramatic swelling of the stellar surface. This,
however, is a transient effect because an important part of the
luminosity is due to accretion, which is reduced by the increase in
the stellar radius.  Only at masses around $300\sm$ does the internal
luminosity become very close to the Eddington value, leading to
runaway expansion of the star and, presumably, the end of accretion.
Omukai \& Palla (2003) considered a range of accretion rates. They
found that if the accretion rate is smaller than
$4.4\times10^{-3}\smyr$, then the total luminosity remains
sub-Eddington and the star continues to grow along the main sequence
to arbitrarily large masses. On the other hand, if the accretion rate
is somewhat larger than this critical rate, the Eddington limit
becomes important at around $100\sm$. Accretion at a rate based on the
settling motions in the core of ABN is slow enough that the Eddington
limit does not affect the final mass. However, these models ignored
the influence of other protostellar feedback mechanisms on the
infalling envelope. These models also assumed spherical symmetry, which
leads to much larger photospheric radii and thus a softer radiation
field than in the more realistic case of disk accretion (see \S 7 in
Paper I).

        Omukai \& Inutsuka (2002) considered the the combined effects
of photoionization heating and continuum radiation pressure due to
photoionization in the infalling envelope.  They show that there is a
critical stellar mass at which the hydrogen ionizing luminosity is
sufficient to create an \ion{H}{2} region, which rapidly spreads to
large distances where its thermal pressure becomes dynamically
important in slowing infall. However, the ionizing radiation force
decelerates the inflowing gas, raising the gas density and therefore
reducing the radius of the \ion{H}{2} region. For spherical inflow,
this mechanism is so effective that the radius of the \ion{H}{2}
region remains well below the gravitational radius $r_g$, stopping any
mass loss.  They concluded that with this effect, there was no limit
on protostellar masses below $1000\sm$. Without this radiation force,
they predicted a mass limit of order $300\sm$. As we shall see below,
these conclusions are sensitive to the assumption of spherical
accretion.

        In Paper I (Tan \& McKee 2004) we modeled the growth of a
primordial protostar from very small masses to large. We included the
effects of rotation of the infalling gas, which led to the formation
of an accretion disk around the protostar.  The goal of this paper is
to determine when the energy output from the protostar is sufficient
to halt accretion and set the final stellar mass.  This is an
extremely complicated problem, the full solution of which requires
three dimensional hydrodynamical simulations that include the
generation, propagation, and dynamical influence of
radiation. Furthermore these simulations must resolve a large range of
scales: the protostar is of order $10\:R_\odot$, while the size of the
quasi-hydrostatic core that encloses $\sim 1000\sm$ is of order 1~pc,
several million times greater. The demands on the time scale are even
greater: the simulation may have to follow the evolution of the star
over its lifetime $\sim 2 - 3$ Myr (Schaerer 2002) while at the same
time following the dynamics of an accretion disk with a characteristic
time scale as short as $10^4$ s. The numerical simulations of ABN were
able to resolve an even greater range of radii, but it will be some
years before it is possible to meet the required dynamical range in
time scales. As a result, we shall develop simple analytic models for
the feedback interaction that we hope will provide a useful first
step.

\begin{figure}[h]
\begin{center}
\epsfig{
        file=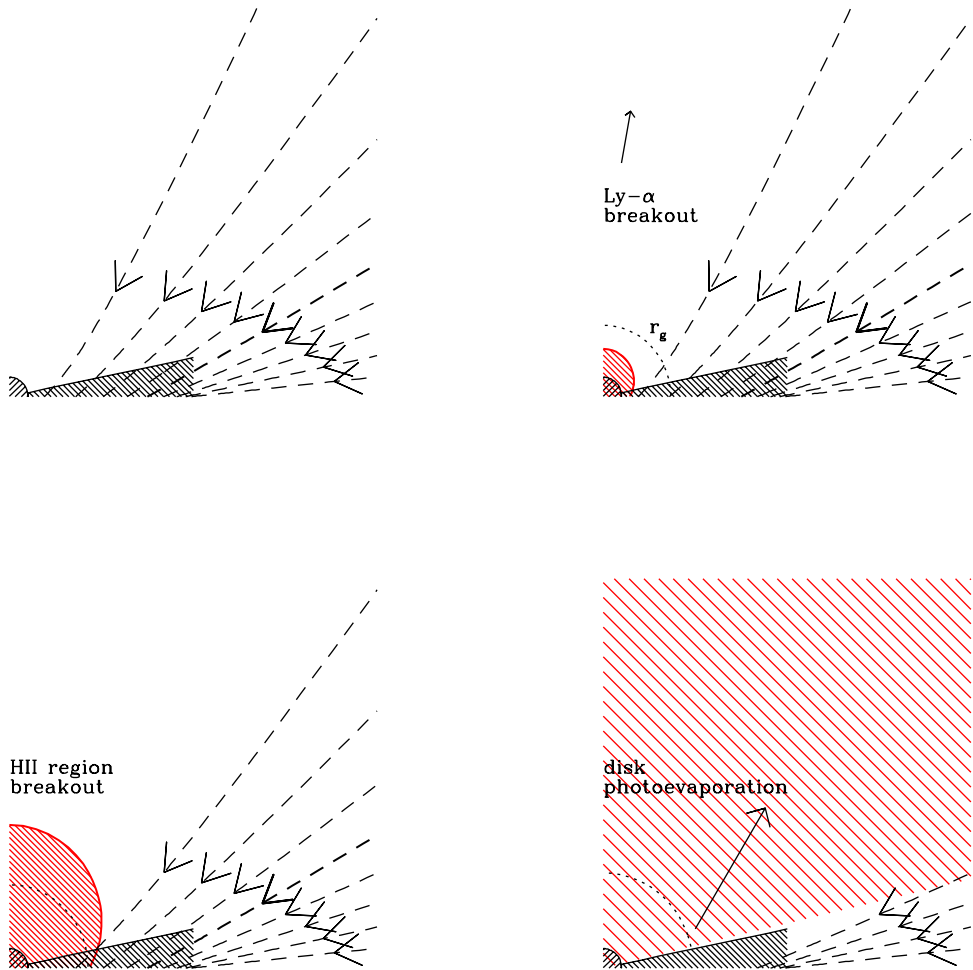,
        angle=0,
        width=6in}
\end{center}
\caption{
\label{fig:overview}
Overview of the accretion geometry and feedback processes involved in
primordial star formation. (a) Top left: Cross section of the
accretion geometry: the dashed lines show streamlines of the rotating,
infalling gas, with figure of revolution from each streamline
separating 10\% of the total infall from this hemisphere. The aspect
ratio of the accretion disk is realistic, while the size of the star
has been exaggerated for clarity. (b) Top right: The shaded region
around the star shows the extent of the \ion{H}{2} region, which at
this relatively early stage is still confined inside the gravitational
radius for the escape of ionized gas, $r_g$. \lal\ radiation pressure
feedback should be strong enough to prevent accretion in the polar
directions. (c) Bottom left: The stellar mass and ionizing luminosity
have grown, and the \ion{H}{2} region is just in the process of
breaking out of the accretion flow. Once a significant volume beyond
$r_g$ is ionized, accretion from these directions is expected to be
shut off. (d) Bottom right: Final stage of accretion involves
shadowing of the equatorial region from stellar ionizing flux by the disk,
which at the same time is photoevaporated. The competition between
this photoevaporative outflow and the residual accretion rate sets the
final mass of the star.}
\end{figure}

        We begin our discussion with a review the results of Paper I
in \S \ref{S:review}.  Feedback effects are then considered in the
approximate order in which they become manifest. In \S\ref{S:dissoc},
we briefly discuss the effects of photodissociation.  \lal\ radiation
pressure feedback, is considered in \S\ref{S:lya} and in several
Appendixes.  The feedback from ionizing photons that can create an
\ion{H}{2} region is considered in \S\ref{S:ion}.  After discussing
shadowing by accretion disks in \S\ref{S:shadow} and an Appendix, the
feedback due to disk photoevaporation is considered in
\S\ref{S:diskevap}.  Figure \ref{fig:overview} gives an overview of
these feedback processes occurring near the protostar.  Finally, our
conclusions are summarized in \S\ref{S:conclusions}.

\section{Review of Paper I: Properties and Evolution of 
Primordial Protostars}
\label{S:review}

The radiative output from a protostar depends on the temperature and
luminosity of its emitting components, which are the star itself
(stellar photosphere), the boundary layer of the accretion disk with
the star, and the larger scale accretion disk. The luminosity of the
star depends mostly on its mass. The size of the star then determines
its temperature. The size of the star and the accretion rate determine
the radiative properties of the both the boundary layer and the
accretion disk, since emission from the latter is dominated by the
inner regions.

The size of the star depends on the accretion rate during its
formation history.  At lower masses there is a balance in the size set
by the need to radiate the luminosity, which is mostly due to
accretion, with a photospheric temperature that is likely to be close
to $\sim 6000$~K because the opacity due to $H^-$ rapidly declines in
this temperature regime.  Under the assumption of spherical accretion,
Stahler, Palla, \& Salpeter (1986) found the protostellar radius to be
\beq
\label{stahler}
r_* \simeq 90.8 \mst^{0.27} \mdst^{0.41}~~~ R_\odot
~~~~~(\mst\la 0.1),
\eeq 
where $\mst\equiv m_*/(100\sm)$
and $\mdsdt\equiv \mdsd/(10^{-3}\;\smyr)$.
For the accretion rates typical of primordial star formation we see that
the size is very large.  For more massive protostars there is a
transition once the star is about as old as its local Kelvin-Helmholtz
time, then contraction proceeds towards the main sequence size, where
accretion can continue. 
In Paper I, we found that for
typical conditions, the protostar reached its main sequence radius
at about $100\sm$. According to Schaerer (2002), this radius
is
\beq
r_*\simeq 4.3\mst^{0.55}~R_\odot~~~~~({\rm main\ sequence})
\eeq
to within 6\% for $0.4\leq\mst\leq 3$.

        Thus almost all the radiative stellar properties depend on
just two parameters: the stellar mass and the accretion rate.  Note
that in principle these properties also depend on the angular momentum
of the infalling gas, since if there was no rotation, then spherical
accretion implies very high gas densities near the protostar,
affecting the location of its photosphere.  However, for any realistic
amount of angular momentum, a disk forms whose size is much larger
than $r_*$, and the star's properties no longer depend on the rotation
of the core from which the star forms.

        The accretion rate of Population III protostars depends on the
density structure of the gas core at point when the star starts to
form. This density structure is set by the balance of thermal pressure
and self-gravity, which in turn depends primarily on the cooling
properties of molecular hydrogen. This cooling creates almost
isothermal cores at $T\simeq 200$~K with an outer bounding density of
about $10^4\;{\rm cm^{-3}}$, which is the critical density of $\rm
H_2$ cooling transitions (for $\rm H_2$ molecules interacting with
atomic H).  In fact the temperatures increase to several hundred K in
the inner parts of the core because of the reduced cooling efficiency
above the critical density. These basic features have been confirmed
by numerical studies (ABN; BCL02; O'Shea \& Norman 2007).  The trigger
for dynamical collapse is thought to be the rapid formation of $\rm
H_2$ by three-body collisions at high densities $\sim 10^{10}$~cm\eee,
since this then dramatically increases the cooling rate in this
region.

        ABN carried their calculations almost to the point of
protostar formation, and at this time gas was flowing inward
subsonically almost everywhere (except for $0.1 M_\odot \la M \la 1
M_\odot$, where the inflow was slightly supersonic).  Shu's (1977)
expansion wave solutions for protostellar accretion are based on the
assumption that the inflow velocity at this time is zero. Hunter
(1977) generalized these solutions and showed that there is a discrete
set of self-similar solutions that begin at rest at $t=-\infty$ and
have a constant infall velocity at the time of protostar formation
($t=0$).  One of these solutions, the Larson-Penston solution (Larson
1969; Penston 1969), has supersonic inflow (Mach number = 3.3 at
$t=0$; this solution is clearly inconsistent with the numerical
results.  In fact, the accretion flow appears to be a settling
solution regulated by \htwo\ cooling. Only one of Hunter's solutions
corresponds to mildly subsonic inflow (Mach number =0.295 at $t=0$), comparable
to that found by ABN, and this is the solution adopted in Paper
I. This solution has a density that is 1.189 times greater than a
singular isothermal sphere at $t=0$, and the accretion rate is 2.6
times greater.

        Hunter's (1977) solution applies to an isothermal
gas. Omukai \& Nishi (1998) and Ripamonti et al. (2002)
have numerically calculated accretion rates
for primordial protostars,
and showed that the accreting gas is isentropic with an
adiabatic index $\gamma\simeq 1.1$ due
to H$_2$ cooling; i.e., each mass element
satisfies the relation $P=K\rho^\gamma$ with
the ``entropy parameter'' $K=$~const. 
In hydrostatic equilibrium, such a gas settles
into a polytropic configuration,
which in general has $P(r)=K_p\rho(r)^{\gamma_p}$.
For an isentropic gas, we have $\gamma_p=\gamma$ and $K_p=K$.
In Paper I, we presented an analytic model for
the protostellar accretion rate for isentropic
gas. We allowed for the existence of an accretion disk
around the protostar with  a significant fraction of the stellar mass,
\beq
\msd=m_*+m_d\equiv (1+f_d)m_*,
\label{eq:fd}
\eeq
with a fiducial value for the disk mass fraction $f_d=\frac 13$.
Following McKee \& Tan (2002, 2003), we wrote the accretion rate as
\beq
\mdsd=\phi_*\frac{\msd}{\tff},
\eeq
where $\phi_*$ is a numerical constant of order unity and
$\tff =(3\pi/32 G\rho)^{1/2}$ 
is the free-fall time of the gas
(measured at $t=0$). 
For gas that is in hydrostatic equilibrium at $t=0$,
McKee \& Tan (2002) showed
that $\phi_*\simeq 1.62-0.96/(2-\gamma_p)$ to within
about 1\% for $0<\gamma_p\leq 1$; we have since verified
that this is valid for $\gamma_p\la 1.2$.
To our knowledge, Hunter's self-similar
solutions starting at $t=-\infty$ have not
been generalized to the non-isothermal
case. 
\footnote{
Fatuzzo, Adams, \& Myers (2004) have given a comprehensive
discussion of self-similar accretion solutions that start
at $t=0$, allowing for inflow velocity, overdensity relative
to hydrostatic equilibrium, and non-isentropic gas
($\gamma\neq\gamma_p$). 
Although they do not treat the time prior to
protostar formation, their isothermal results for $t>0$ are
consistent with Hunter's, as expected.
For the non-isothermal case, Fatuzzo et al. (2004) present
results for gas that is inflowing at $r\rightarrow\infty$, but
these are not self-similar in that the accretion rate 
depends on where the integration begins (F. Adams, private
communication). However, it is possible to generalize their treatment
so that the Mach number of the inflow is constant.
Presumably the overdensity
and infall Mach number of the $\gamma=1.1$ analog
of the mildly subsonic
Hunter solution are similar to those of the isothermal
solution; if they are exactly the same, then
the accretion rate would be about 2.0 times
that for the case of hydrostatic initial conditions,
somewhat less than the isothermal value of 2.6.}
In Paper I, we therefore assumed that
the accretion rate for the $\gamma=1$ case
is 2.6 times greater than
that for hydrostatic initial conditions, just
as in the isothermal case.

Feedback from the star, whether due to winds, photoionization,
or radiation pressure, can reduce the accretion
rate onto the star.
We define a hypothetical
star-disk mass, $\msdo$, and accretion rate, $\mdsdo$,
in the absence of feedback. In this case, the star-disk
mass equals the mass of the core out of which it
was formed, $\msdo=M(r)$.
The instantaneous
and mean star formation efficiencies are
\beqa
\esd&\equiv&\frac{\mdsd}{\mdsdo}, 
\label{eq:esd}\\
\esdb&\equiv&\frac{\msd}{\msdo}=\frac{\msd}{M}.
\label{eq:esdb}
\eeqa
In our previous work, we assumed that the star formation efficiency was
constant, so that $\esd=\esdb$. in the present work, we shall find that
significant feedback does not set in until the star is fairly massive, so
that we must distinguish the instantaneous and mean values.
In Paper I, we found that the accretion rate onto the star-disk system 
is
\beq
\mdsd=0.026 \left[\frac{\epsilon_{*d}K'^{15/7}}
        {(M/\sm)^{3/7}}\right]~~~~~M_\odot~{\rm yr}^{-1},
\eeq
where
\beq
K'\equiv \frac{P/\rho^\gamma}{1.88\times 10^{12}~{\rm cgs}}=
         \left(\frac{\teff'}{300~{\rm K}}\right)\left(\frac{10^4~
         {\rm cm}^{-3}}{n_{\rm H}}\right)^{0.1}
\eeq
is a measure of the entropy of the accreting gas. Here $\teff'
\equiv T+\mu\sigma^2_{\rm turb}/k$ is an effective temperature
that includes the effect of turbulent motions; we have added
a prime to the $\teff$ defined in Paper I to distinguish it
from the effective temperature of a radiating atmosphere.
Expressing the accretion rate in terms of the stellar mass,
which equations (\ref{eq:fd}) and (\ref{eq:esdb}) imply is
$m_*=\esdb M/(1+f_d)$, we find
\beq
\mdsd=0.026 \left[\frac{\epsilon_{*d}\esdb^{3/7}K'^{15/7}}
        {(1+f_d)^{3/7}(m_*/\sm)^{3/7}}\right]~~~~~M_\odot~{\rm yr}^{-1},
        \label{eq:mdsd}
\eeq
With $K'=\esd=\esdb=1$, 
this result is in good agreement with the results
of Omukai \& Nishi (1998) and Ripamonti et al. (2002);
since their 1D calculations did not allow for disks,
this comparison is made for $f_d=0$.
Note that this agreement validates our use of the Hunter
mildly subsonic solution to infer the accretion rate.
If $\epsilon_{*d}=1$ (i.e., no feedback) and $f_d=\frac 13$  
then the accretion rate onto the star $+$ disk is
\beq
\mdsdt\rightarrow 3.20\left(\frac{K'^{15/7}}{\mst^{3/7}}\right),
\label{eq:mdst}
\eeq
where henceforth it will be understood that numerical evaluations
denoted by ``$\rightarrow$'' have $\esd=1$ and $f_d=\frac 13$.
In this case the accretion rate onto the star (which may be primarily
through the disk) is 3/4 of this [since $\mds=\mdsd/(1+f_d)$].

Our estimate of the accretion rate is somewhat above that estimated by
ABN, but this is to be expected since their calculation stopped prior
to the formation of the protostar.  Indeed, at the time at which the
protostar first forms ($t=0$), the accretion rate at any finite radius
$r$ [i.e. $\dot m(r)=4\pi r^2 \rho |v_r|$] in a self-similar,
isothermal collapse is smaller than the value it has at times $t >
r/c_s$, where $c_s$ is the isothermal sound speed.  Equivalently, at a
given time, the accretion rate at radii $r\ga c_s t$ is less than that
at small radii, $r\ll c_s t$.  In Shu's (1977) solution for the
collapse of a singular isothermal sphere, the accretion rate at a
given time is zero outside the expansion wave at $r=c_s t$; inside the
expansion wave, the accretion rate smoothly increases to $0.975
c_s^3/G$ as $r\rightarrow 0$. For the Larson-Penston solution, the
accretion rate at a given time $t>0$ increases from $29 c_s^3/G$ at
large radii ($r\gg c_s t$) to $47 c_s^3/G$ at small radii ($r\ll c_s
t$). For Hunter's mildly subsonic solution, which we have suggested is
closest to the numerical simulations, the accretion rate increases
from $0.70 c_s^3/G$ at large radii to $2.58c_s^3/G$ at small radii
(Hunter 1977), an increase of a factor 3.7. This demonstrates that
caution should be exercised in inferring accretion rates at late times
from those measured at early times, which is commonly done in
simulations (e.g., ABN, Yoshida et al. 2006, O'Shea and Norman 2007).

        The age of the star when it reaches a mass $m_*$ is (Paper I)
\beq
\tyr=27\left(\frac{1+f_d}{\esdb}\right)^{10/7}K'^{-15/7}\left(\frac{m_*}{M_\odot}\right)^{10/7}
        \rightarrow 2.92\times 10^4 K'^{-15/7}\mst^{10/7},
\eeq
where $\tyr\equiv t/(1$~yr)
and where it is the mean star formation efficiency $\esdb$ that enters.
The resulting stellar mass is
\beq
m_*\rightarrow 0.075 K'^{1.5} \tyr^{0.7}~~\sm.
\eeq
Bromm \& Loeb (2004) have carried out a 3D simulation of the
accretion onto the protostar for the first $10^4$ yr, and
for $K'=1$ our result
is within a factor $\sim 2$ of theirs for this time
period. (However, it should be noted that an extrapolation
of their result to times beyond $5\times 10^4$ yr
gives a mass less than half our estimate of the mass of the star plus disk. 
It remains to be determined whether such an extrapolation is valid.)

        With this estimate of the protostellar mass, it is possible to
calculate the maximum possible mass a primordial star could have.
Schaerer (2002) has calculated the main sequence lifetimes of
primordial stars with no mass-loss for $m_*\leq 500\sm$, and his
results are accurately described by the expression $t_{\rm ms}\simeq
2.7\mst^{-0.24}$~Myr for $100\sm\la m_*\la 500 \sm$.  If we assume
that the accretion is not limited by any feedback processes
($\esd=1$), that Schaerer's results can be extrapolated to higher
masses, and that accretion during the relatively short post
main-sequence phase is negligible, then we find 
\beq m_{*,\,{\rm max}}
=\int_0^{t_{\rm ms}}\mds\, dt \simeq
1900\left(\frac{1}{1+f_d}\right)^{0.86}K'^{1.28}~~~\sm\rightarrow 1500
K'^{1.28}~~~\sm.
\label{eq:mmax}
\eeq
The maximum possible stellar mass is thus controlled
by the value of the entropy parameter of the core.

        In Paper I we also considered the effect of rotation.
Rotation of the infalling gas has a dramatic effect
on the evolution of the protostar, since it leads to 
much smaller photospheric radii and correspondingly
higher temperatures and accretion luminosities. 
We parameterized the rotation in terms of
\beq
\fkep\equiv\frac{v_{\rm rot}(r_{\rm sp})}{v_{\rm Kep}(r_{\rm sp})},
\label{eq:fkep}
\eeq
the ratio of the rotational velocity to the Keplerian velocity
measured at the sonic point
at $r_{\rm sp}$. ABN found $\fkep\sim 0.5$ independent
of radius, so we adopt this as a fiducial value. We then showed
that the accreting gas formed a disk with an outer radius
\beqa
r_d &=& 1.92\times 10^{16}\left(\frac{\fkep}{0.5}\right)^2\left(
        \frac{\msdt}{\esdb}\right)^{9/7}K'^{-10/7}~~~~~{\rm cm},
        \label{eq:rd1}\\
        &\rightarrow& 2.78\times 10^{16} \left(\frac{\fkep}{0.5}\right)^2
        \frac{\mst^{9/7}}{K'^{10/7}}~~~~~{\rm cm}.
\label{eq:rd2}
\eeqa

\section{Photodissociation Feedback}
\label{S:dissoc}

\begin{figure}[h]
\begin{center}
\epsfig{
        file=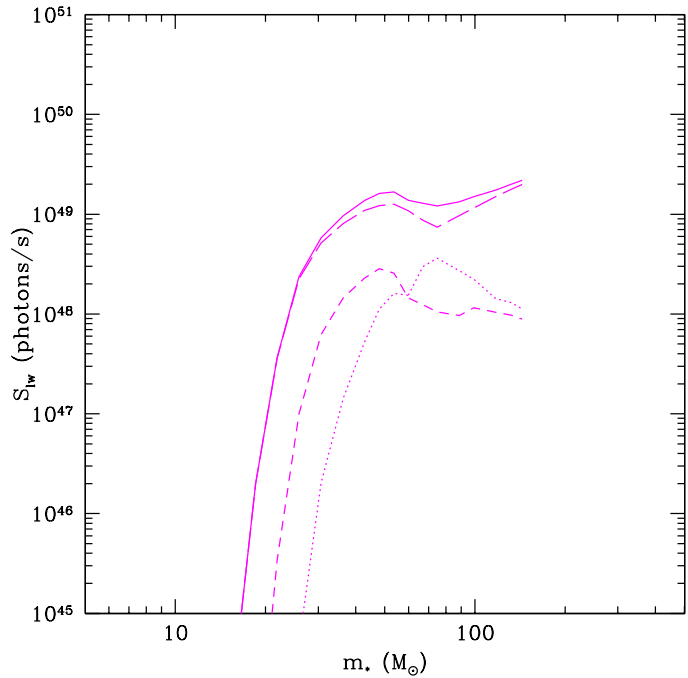,
        angle=0,
        width=\figwidth}
\end{center}
\caption{
\label{fig:Slwstarevol}
Evolution of Lyman-Werner photon luminosity from the fiducial model of
primordial star formation, including effects of stellar feedback. The
total (solid line) and contributions from the protostellar surface
(long-dashed line), boundary layer (dashed line), and accretion disk
from $r < 10r_*$ (dotted line) are shown. }
\end{figure}

        As the protostar grows in mass, it begins to emit copious
amounts of non-ionizing ultraviolet radiation (far-ultraviolet, or
FUV, radiation), as shown in Figure \ref{fig:Slwstarevol}. This
radiation can photodissociate the \htwo\ that is critical for cooling
the accreting gas (Omukai \& Nishi 1999; Glover \& Brand 2001); its
dynamical effects are considered in the next section.

        Once the molecular coolants in the
accreting gas are destroyed,
the adiabatic index rises from
$\gamma\simeq 1.1$ to $\gamma=\frac 53$. If the
gas were able to continue contraction,
it would heat up until the temperature
became high enough ($T\sim 10^4$ K) 
to excite the Ly-$\alpha$ transition. For $T\ga 10^4$ K,
the adiabatic index would then drop back to about 1.

        Can the protostar continue to accrete when
$\gamma\simeq \frac 53$? If one considers the related
problem of the gravitational stability of 
polytropic gas spheres,
one might be led to conclude that accretion
would stop: polytropic stars are stable
against gravitational collapse for $\gamma>\frac 43$
(Chandrasekhar 1939), and the same is true for polytropic
gas clouds even if $\gamma_p <\frac 43$ (McKee
\& Holliman 1999). 
However, there is a crucial distinction between collapse
onto a protostar and the contraction of a gas cloud
prior to protostar formation, 
and that is the presence of the central protostar, which
is effectively a mass singularity. The stability analyses
cited above assumed that there was no mass singularity
at the origin. When one is present, the problem is
analogous to that of Bondi accretion, 
the accretion of non-self gravitating gas onto
a star; this can 
occur for $\gamma=\frac 53$ (e.g., Shapiro \& Teukolsky
1983). The problem of protostellar accretion,
which includes the self-gravity of the gas,
has been considered 
by Fatuzzo et al. (2004) for a wide range
of conditions. They showed
that gravity dominates over pressure close to
the protostar, so that accretion
can occur, provided $\gamma < \frac 53$.
It is straightforward to see why: In supersonic
inflow, the density scales as $\rho\propto
r^{-3/2}$, so that the temperature 
$T\propto r^{-3(\gamma-1)/2}$ rises
more slowly than the kinetic energy per unit
mass $\propto r^{-1}$ provided $\gamma<\frac 53$.
They demonstrated that the accretion rate
for a singular, initially isothermal sphere
with $\gamma=1.6$ is only slightly smaller
than for the case in which $\gamma=1$.

        The argument of Fatuzzo et al. applies
to the inner, supersonic region of infall.
What about the outer, subsonic region?
There the density varies as a higher 
power of radius (e.g., for $v_{\rm infall}$ = const.,
$\rho\propto r^{-2}$), so that pressure
can overcome gravity at a lower value of $\gamma$.
To see this more quantitively, consider the 
case of a singular isothermal sphere with $\gamma\neq 1$.
Assume that the initial density of the sphere is
$\Lambda$ times greater than the equilibrium
value. If the gas is flowing inward at a velocity
$-v_\infty$ far from the protostar (i.e., at
large values of the similarity variable $x$, which
is just $r/[c_s t]$ in the isothermal case), then $v$ varies as
\beq
v=-v_\infty-\frac{2(\Lambda-1)}{x}-\frac{(4-3\gamma)v_\infty}{x^2}
        +\cdots\ .
\eeq
(Fatuzzo et al. 2004; we have corrected a typo in the last factor).
Thus, for $\gamma>\frac 43$, pressure forces will tend
to decelerate the flow; however, this can be overcome by a suitable
overdensity $\Lambda$. We have confirmed this by 
numerical integration of the equations given by Fatuzzo 
et al.: For $v_\infty>0$ and $\gamma\leq \frac 43$,
accretion is possible for $\Lambda\geq 1$; for
$\frac 53 \geq\gamma>\frac 43$, accretion is possible
provided $\Lambda$ exceeds some threshold.
For primordial star formation, we estimate
$\gamma_p=1.1$ and $v_\infty/c_s\simeq 0.3-0.5$;
accretion can occur in this case for $\gamma=\frac 53$ for $\Lambda
>1.16,\ 1.28$, respectively. These overdensities are quite
modest (for example, Hunter's 1977 subsonic infall
solution has $v_\infty/c_s=0.295$ and $\Lambda=1.189$), 
so we anticipate that photodissociation should
not prevent protostellar accretion. The value of 
the overdensity is likely to vary from one protostar
to another, however, so it is possible that
in some cases it would be too small to
permit accretion. In such cases
the infalling gas would decelerate; once it is
stationary, however, it could resume accretion
when it is overtaken by an expansion wave,
as shown by Fatuzzo et al. 
Our numerical calculations show that the increase in
$\gamma$ from 1.1 to $\frac 53$
has only a minor effect on the accretion
rate, diminishing it by less than 20\%.
We conclude that photodissociation of
molecular coolants by the protostar
does not have a significant effect on
its accretion rate.

On the other hand, collapsing cores that do not contain a protostar,
but that are close enough to a protostar that their molecular coolants
are destroyed, will cease collapsing if their central temperature is
low enough ($<10^4$~K) that $\gamma$ exceeds $\frac 43$. Thus, FUV
emission from the first stars is potentially quite effective at
suppressing star formation in their vicinity.  We can estimate the
distance over which star formation is suppressed from the work of
Glover \& Brand (2001). As in Paper I, we assume that the core is in
approximate hydrostatic equilibrium and is characterized by an entropy
parameter $K$.  We find that the time to dissociate \htwo\ is less
than the free-fall time $\tff$ if the core is within a distance \beq
D=24\left[\left(\frac{S_{\rm LW}}{10^{49}\ \s^{-1}}\right)
\left(\frac{10^{-3}}{x_2}\right) \left(\frac{f_{\rm abs}f_{\rm
diss}}{0.01}\right)\right]^{1/2} \frac{1}{\bar
n_4^{21/40}K'^{1/4}}~~~~~{\rm pc}, \eeq of the protostar, where
$S_{\rm LW}$ is the photon luminosity in the Lyman-Werner bands, $x_2$
is the fractional abundance of \htwo, $f_{\rm abs}$ is the fraction of
the Lyman-Werner flux absorbed by the \htwo, $f_{\rm diss}$ is the
fraction of absorptions that result in dissociation, and $\bar n$ is
the mean density of H nuclei.  Thus, even a $100 \sm$ star, which has
$S_{\rm LW}\simeq 10^{49}$~s\e, can suppress star formation in an
existing core only if the core is relatively nearby.  A more detailed
analysis by Susa (2007) comes to the same conclusion.  Ahn \& Shapiro
(2007) model both dissociation and ionizing feedback and also find a
relatively weak suppression of PopIII.2 star formation by neighboring
PopIII.1 stars. Whalen et al. (2008) have presented multi-dimensional
numerical simulations of these processes.

\section{\lal\ Radiation Pressure}\label{S:lya}

        The second feedback effect of FUV radiation is radiation
pressure acting on the Lyman absorption series in the infalling
neutral gas.  This effect has been considered previously by Oh \&
Haiman (2002), who studied feedback effects in halos with virial
temperatures above $10^4$ K, which are more massive than those we
consider. They concluded that \lal\ radiation pressure could be
important, but did not find any constraint on the mass of the star
that could form. Our work improves upon theirs in several respects: we
include stellar continuum photons injected away from line center as
well as \lal\ photons emitted in the \ion{H}{2} region; we allow for
Rayleigh scattering; we include the limitations on the radiation
pressure set by two-photon emission and by the blackbody constraint;
and we allow for the effects of rotation in the infalling gas.

Since conditions are very opaque, the \lal\ radiation can be
considered to be isotropic.  The \lal\ radiation pressure is then \beq
P_\alpha=\frac{1}{3} u_\alpha= \frac{4\pi J_\alpha}{3c}, \eeq where
$u_\alpha$ and $J_\alpha$ are the energy density and mean intensity of
the \lal\ radiation.  The estimation of $J_\alpha$ is complicated by
the fact that \lal\ photons can diffuse in frequency as well as in
space, and that at the optical depths we are considering, the transfer
is dominated by the damping wings of the line profile (Adams 1972).
This problem is far too difficult to treat analytically for the
geometry and dynamics that we are using to model the protostellar
accretion. We therefore make the following substantial approximations
when evaluating $J_\alpha$ at the outer boundary of the \ion{H}{2}
region, $\rtwo$ (which may be at the surface of the protostar), and at
a particular polar angle.  (1) The axially symmetric geometry can be
replaced by an equivalent slab geometry. The effects of spherical
divergence are incorporated by normalizing the mean intensity to the
flux at $\rtwo$.  The slab column is set equal to 20\% of that in the
infalling gas based on the discussion in Appendix \ref{app:radpress}.
(2) The anisotropy in the optical depth can be accounted for by taking
the harmonic mean of the opacity, \beq
\frac{1}{\taueff}=\frac{1}{A}\int \frac{dA}{\tau({\bf r})}
\label{eq:harmonic}
\eeq
(see Appendix \ref{app:aniso}).
In practice, the escape of photons is primarily controlled by the minimum 
optical depth, which is in the polar direction, so in our numerical models we
evaluate $\taueff$ 
with a column density that is 20\% of the column 
in the vertical direction from 
the point of interest. 
(3) Finally, we assume that the effect of the velocity field can be
approximated by a Doppler line profile of suitable width (see below).

   The propagation of resonance line photons in very opaque media has
been treated by a number of authors (Adams 1972;
Harrington 1973; Hummer \& Kunasz 1980; Neufeld 1990).
Let the mean optical depth in the line be
\beq
\bar\tau=\frac{1}{\dnd}\int \tau_\nu d\Delta\nu.
\eeq
In terms of the normalized frequency $x\equiv \Delta\nu/\dnd$, we have
$\tau_x=\tau_\nu=\bar\tau\phi_x$,
where $\phi_x$ is the line profile.
In the Doppler core, the line
profile is $\phi_x\simeq\exp(-x^2)/\surd\pi$; in
the damping wings it is  $\phi_x\simeq a/(\pi x^2)$,
where $a$ is the ratio of the natural line width to the Doppler width. In
applications, we generally have $a\ll 1$, and in that case
the optical depth
at line center, $\tau_0$, is related to mean optical depth by
$\tau_0=\bar\tau\phi_0=\bar\tau/\surd\pi$.
The opacity
is $\kappa_x=\bar\kappa\phi_x$, and the mean free path
is $\ell_x=1/\kappa_x$.

      As shown by Adams (1972),
resonance photons escape in a single longest excursion from line
center. After $n$ scatterings, the escaping photon has a frequency shift
$x_e\simeq n^{1/2}$ and it has traveled a distance
$n^{1/2}\ell_e\simeq n^{1/2}/(\bar\kappa\phi_e)$,
where $\ell_e=\ell(x_e)$, etc.  In
order for the photon to escape, this distance must be the size of
the region, $L=\bar\tau_L/\bar\kappa$.  This implies
$n^{1/2} \simeq \bar\tau_L\phi_e$ and $x_e\simeq\bar\tau_L\phi_e$, which
in turn gives the characteristic frequency of the escaping
photons as $x_e\sim (a\bar\tau_L)^{1/3}$.  The total path
length traversed by the escaping photons is about $n^{1/2}L$.
As a result, we expect the mean intensity to exceed the
incident intensity by a factor of about $n^{1/2}\sim
(a\bar\tau_L)^{1/3}$.

        The velocity field has contributions from thermal motions, turbulent
motions, and the overall flow.  Thermal and microturbulent velocities are
naturally included in $\Delta v_D$.  In the cases we shall
consider, the overall flow is highly opaque, so it generally does not
contribute to the random walk of the photons.  If the velocity width of
the flow $\Delta v_f$ (including macroturbulence) is small compared to
the line width of the escaping photons, $\Delta v_e\simeq
(a\taueff)^{1/3}\Delta v_D$ (from Neufeld 1990), then the flow has
negligible effect on the escape of the photons.  On the other hand, if
$\Delta v_f\gg\Delta v_e$, then the effective column density of the gas
is reduced.  For example, in the simple case in which the velocity varies
linearly with the column density, photons will interact with only a
fraction $\Delta v_e/\Delta v_f$ of the gas.  If $\Delta
v_{e0}$ is the line width in the absence of any flow velocity, then one
can show that the effective column density is reduced by a factor of
about $(\Delta v_{e0}/\Delta v_f)^{3/2}$ from the total value.
In our numerical models we always set $\Delta v_D=12.9\kms$, the value
appropriate for $T=10^4\:{\rm K}$ gas,
the assumed temperature of the infalling neutral
gas near the protostar.
We set $\Delta v_f$ equal to
half the difference in radial velocities of the inner and outer edges
of the slab. If $\Delta v_{e0} > \Delta v_f$, which is not usually the
case, we reduce the effective column by the factor $(\Delta
v_{e0}/\Delta v_f)^{3/2}$.

        Appendix \ref{app:enhance} describes the general enhancement
in the intensity of photons that are trapped by the \lal\ damping
wings, and, if the columns are very large, by the opacity due to
Rayleigh scattering.  Thus, photons from the protostellar continuum,
outside the frequency interval defined by $x_e$, can contribute to the
radiation pressure.  The enhancement in intensity leads to an increase
in the radiation pressure so that the momentum transferred from the
radiation to the gas is $F/c$ in each optical depth.  As shown in
Appendix \ref{app:enhance}, the isotropic component of the radiation
pressure is
\beq
\pradi=\frac{4\pi J_{\rm iso}}{3}=\frac{4\pi}{3} 
        \int d\nu_i\;{\rm Min}\left\{B_{\nu_i},\;
        \frac{8.25 N_{\rm eff,20}^{1/3} \dvds^{-2/3}(F_{\nu_i}/c)}
        {{\rm Min}[1,\;2.62N_{\rm eff,20}^{1/3} \dvds^{-2/3}]
        +5.41[\hat{x}_i^2/f(\nu_i)] +\Gamma(\hat{x}_i)}\right\},
\label{eq:pfinal}
\eeq
where $N_{\rm eff,\, 20}\equiv N_{\rm eff}/(10^{20}\:{\rm cm^{-2}})$
and $B_{\nu_i}$ is the intensity of a blackbody with a
temperature equal to that at the protostellar surface. When this limit
is evaluated for the reprocessed \lal\ photons, the intensity is
limited to that of a blackbody at the temperature of the ionized gas
in the \ion{H}{2} region (see Appendix \ref{app:enhancebb}).
This expression is valid provided $\taueff\ga 1/a$, corresponding
to $N_{\rm eff}\ga 10^{16}\dvds^2$~cm\ee\ for Lyman $\alpha$.

        What is the condition for the radiation pressure to
halt the accretion?  We assume that the accreting gas
is inside the sonic point, so that the gas pressure
is negligible.
For steady, radial flow, the equation of motion of the gas is 
\beq
\rho v\,\frac{dv}{dr}=-\frac{d\pradi}{dr}-\frac{\rho Gm_*}{r^2},
\eeq
since we have shown in Appendix \ref{app:radpress} that the radiative
force can be represented by the gradient of the isotropic
component of the radiation pressure.
We assume that
the radiation pressure builds up rapidly over a distance small
compared to the radius $r$; this is justified below.  Then 
constancy of the mass flux implies $\rho v\simeq$~const. 
If the radiative force is to stop the flow in a small distance, then the
gravitational force must be negligible in comparison.  We then have
\beq
\frac{d}{dr}\;\left(\rho v^2+\pradi\right)\simeq 0,
\eeq
so that $\rho v^2 + \pradi\simeq$~const in the deceleration
region. When gas enters this region, 
the radiation pressure is small
and $v\simeq v_{\rm ff}$, the free-fall velocity; as the gas decelerates,
the radiation pressure rises and $v$ drops.
The inflow will be halted if the radiation
pressure at the inner edge of the
deceleration region is $\pradi=\rho_{\rm ff} v_{\rm ff}^2$,
where $\rho_{\rm ff}$ is the density in the freely falling 
gas.\footnote{Jijina \& Adams (1996) have given an alternative
criterion based on treating the radiative force per unit mass as the
gradient of a potential. Their approach is appropriate 
when one knows the spatial variation of the force in advance,
which is not the case here. One can show that the two approaches
are equivalent if the radiative force per unit mass falls
off rapidly with $r$.}
We have verified this
simple argument by solving for the flow in the case that
the flux varies as $r^{-k_F}$, with $k_F>2$; such a faster than
spherical fall-off in the flux is expected when the density
distribution is not spherically symmetric, so that flux will
escape into regions of lower opacity 
(as in the case of the ``flashlight effect''---Yorke and Bodenheimer
1999).  We find that the radiation pressure required
to halt the infall is
within about 10\% of $\rho_{\rm ff} v_{\rm ff}^2$ for $k_F>2.5.$
In order to reverse the
inflow and eject the matter, the radiation pressure must be twice
this, $\pradi=2 \rho v_{\rm ff}^2$. Of course, 
a steady radial flow cannot reverse direction. To see that the
factor 2 is required, one can imagine that the flow is inward
over  half the sky and outward over the other half; to maintain
the same accretion rate, the density would have to be twice as
large.
If the gas is initially in 
a disk, there is no infall to start with
and the pressure required for ejection is $\rho 
v_{\rm Kep}^2=\rho v_{\rm ff}^2/2$.

\begin{figure}[h]
\begin{center}
\epsfig{
        file=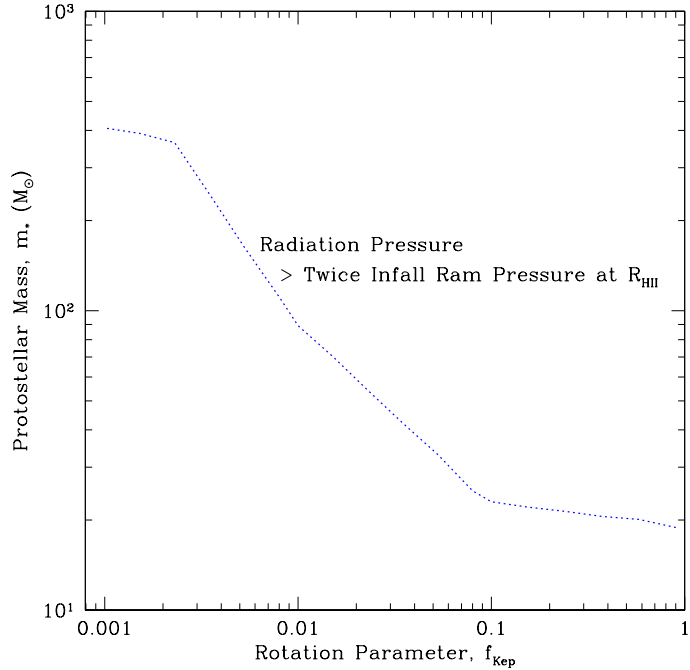,
        angle=0,
        width=\figwidth}
\end{center}
\caption{
  \label{fig:mbreaklal}
Protostellar mass scale at which \lal\ radiation pressure becomes
twice the ram pressure of the infalling gas at the edge of the
\ion{H}{2} region along the polar direction as a function of $f_{\rm
Kep}$. At this point the radiation pressure is expected to reverse the
accretion in the polar direction and evacuate a polar cavity, through
which \lal\ photons can escape. Thus this feedback mechanism will act
to reduce the efficiency of accretion, but will not significantly
impede the growth of the star, since most mass is accreted from directions away from the polar axis.}
\end{figure}

We evaluate this criterion along the polar axis at the edge of the
\ion{H}{2} region, which is where the ram pressure of infalling gas is
a minimum and where breakout should occur first (Figure
\ref{fig:mbreaklal}). At low values of $f_{\rm Kep}$ the breakout does
not occur until the star has reached several hundred solar masses as
the photosphere is very large and cool, producing little FUV flux. At
reasonable values of $f_{\rm Kep}\gtrsim 0.1$, polar breakout can
occur relatively early, at $\sim 20\:M_\odot$. This is the point in
the protostellar evolution when the star is starting its rapid
contraction to the main sequence, and the surface temperature and
luminosity are thus rising.  For these values of $f_{\rm Kep}$ the
densities and ram pressures become significantly greater as the sight
line moves away from the pole. By the time that the radiative flux
from the star is large enough to reverse the flow in these directions,
a polar cavity would have been blown out, thus reducing the
enhancement in the radiation pressure due to trapping of photons. Thus
although \lal\ radiation pressure can act to reduce the efficiency of
accretion, we expect it to be unable to stop it. Even the reduction
of the accretion efficiency is likely to be relatively modest, since
even a small polar cavity can dramatically reduce the radiation
pressure in the \ion{H}{2} region. In the following sections we
consider other feedback mechanisms that are more effective at limiting
accretion, although they do so at higher masses. When following the
stellar evolution to these masses we shall assume the reduction in
accretion efficiency due to \lal\ feedback is negligible.

\section{Ionizing Feedback and Breakout of the \ion{H}{2} Region}
\label{S:ion}

\subsection{Photoionization Heating}

Extreme ultraviolet ($h\nu>13.6$~eV) radiation from the protostar can
ionize infalling neutral gas, creating an \ion{H}{2} region. The
temperature of the 
ionized gas is about
$\sim 2.5\times 10^4$~K, 
based on the models of Giroux \& Shapiro (1996) 
and Shapiro, Iliev and Raga (2004) with stellar spectra.
The thermal pressure $P\equiv \rho c^2$ of the ionized
region is typically much greater than that in neutral gas of the same
density because of the elevated temperatures and sound speeds: $c_i =
11.6 (T_i/10^4\:{\rm K})^{1/2}\kms$. The pressure gradients created at
this ionized-neutral boundary can become steep enough to 
cause the \ion{H}{2} region to expand and to dramatically
reduce the
accretion of gas to the star. In this section we attempt to calculate
the point in the protostellar evolution at which this transition
occurs. 
This problem has been considered previously by Omukai
\& Inutsuka (2002). The new feature in our treatment is that
we allow for the rotation of the infalling gas, which can
significantly reduce the density near the protostar.
As we shall see, this completely changes the evolution
of the \ion{H}{2} region.

        As in Paper I,
we approximate the density distribution of the infalling gas
by the Ulrich (1976) solution.
The gas is assumed to be spherically symmetric 
far from the protostar, and
each mass element conserves its angular
momentum as it falls ballistically toward the star.
Terebey, Shu, \& Cassen (1984) showed that this solution
matches on to an expansion-wave solution for the gravitational
collapse of a singular isothermal sphere.
The resulting density distribution is
\beq
\rho=\frac{\mdsd \psi(\mu,r)}{4\pi r^{3/2}(2G m)^{1/2}},
\label{eq:rho}
\eeq
where $\mu=\cos\theta$, 
\beq
\psi(\mu,r)= \left(\frac{2}{\displaystyle 1+\frac{\mu}{\mu_0}}
        \right)^{1/2}\frac{1}
        {\displaystyle \frac{\mu}{\mu_0}+2\mu_0^2
        \left(\frac{r_d}{r}\right)},
\label{eq:psi}
\eeq
and $\mu_0$ is the value of $\mu$ far from the protostar.
The two angles are related by
\beq
\frac{r_d}{r}=\frac{\mu_0-\mu}{\mu_0(1-\mu_0^2)},
\label{eq:rdr}
\eeq
which shows that $\mu_0>\mu$: the gas converges
toward the disk plane. 
Ulrich assumed that the disk had negligible mass,
so that $m=m_*$ in equation (\ref{eq:rho}). In our case,
$m$ varies smoothly from $\msd$ to $m_*$ as $r$ shrinks
from being much greater than $r_d$ to being much less
than $r_d$. This variation in the mass acting on the
infalling gas leads to small, unknown deviations from
the Ulrich solution. In view of the necessarily
approximate nature of the solution and the fact that
$\msd$ and $m_*$ differ by only a factor $\frac 43$ in
the fiducial case, we shall set $m=\msd$ in applying
equation (\ref{eq:rho}).

        The density factor
$\psi$ depends on both the current direction cosine, $\mu$, as
well as the initial one, $\mu_0$, with the two being related
by the cubic equation (\ref{eq:rdr}). 
In our analysis, it is convenient to have an 
approximation for $\psi$ in which the dependence on $\mu_0$
is eliminated:
\beq
\psi\simeq \left[\frac{2}{1+{\rm Max}(\mu^{2/3},1-\zeta)}\right]^{1/2}
        \frac{1}{\displaystyle 0.5\left(\zeta-1+3|\zeta-1|\right)+
        3\mu^{2/3}{\rm Min(1,\zeta)}},
\label{eq:psia}
\eeq
where $\zeta\equiv r_d/r$.
This is exact for all $r$ at $\theta=0$, where $\mu=\mu_0=1$, and
at $\theta=\pi/2$, 
in the plane of the disk. 
It is also exact at $r=r_d$ for all $\theta$.
For $r<r_d$ it is accurate
to better than 20\%;
for $r>r_d$, it is accurate to within a factor 2.
To simplify our results, we
approximate this further for $r\la 0.5 r_d$ and take
\beq
\psi\sim \left(\frac{2}{1+\mu}\right)^{1/2}\frac{r}{2r_d},
~~~~~(r\la 0.5 r_d).
\label{eq:psiaa}
\eeq
This approximation is quite accurate at
$\mu=\frac 23$ (better than 20\% for $r\leq r_d$), but
it deteriorates away from there, underestimating
the density by a factor $\sim 2$ in the plane for $r=\frac 12 r_d$
(the accuracy improves as $r$ decreases).
Nonetheless, it suffices to give an analytic estimate
for the behavior of the \ion{H}{2} region.

\subsubsection{Early Evolution of the \ion{H}{2} Region}

        \ion{H}{2} regions are bounded by ionization fronts.
Ionization fronts that move faster than about $2c_i$ with respect
to the neutral gas,
where $c_i$ is the isothermal sound speed of
the ionized gas, are termed ``R-type;'' such fronts have
little dynamical effect (Spitzer 1978). However, if the
velocity of the front slows below $2c_i$, a shock forms in
front of the ionization front and the velocity of
the front into the shocked gas falls to $\simeq c_n^2/2c_i$, where
$c_n\ll c_i$ is the isothermal sound speed of the shocked neutral
gas; such ionization fronts are termed ``D-type.''
When the \ion{H}{2} region first forms, it is embedded in
gas falling inward with a velocity $\vff\gg 2c_i$. As
a result, the ionization front is initially R-type, and
the radius of the \ion{H}{2} region, $\rtwo$,
is determined by ionization balance.

        Since the density of the infalling gas depends on the
angle $\theta$ relative to the axis of rotation,
$\rtwo$ depends on angle also.
We determine this radius in the sector approximation,
in which ionizations balance recombinations in each element
of solid angle:
\beq
\frac{dS}{d\Omega}=\frac{S}{4\pi}
        =\int_{r_*}^{\rtwo} r^2\atwo n_e n_p dr,
\label{eq:somega}
\eeq where $S$ is the rate of emission of ionizing photons and $\atwo
\simeq 2.6\times 10^{-13}T_4^{-0.8}$~cm$^3$~s\e\ is the recombination
rate to the excited states of ionized hydrogen.  In writing equation
(\ref{eq:somega}), we have made three approximations. First, we have
assumed that the rate of emission of ionizing photons is much greater
than the rate of accretion of hydrogen atoms so that ionizations and
recombinations are very nearly in balance 
(note that advection of neutral H into the \ion{H}{2} region is allowed
for in the numerical models).
For a mass accretion rate of $10^{-3}\ M_\odot$ yr\e, the
hydrogen accretion rate is about $3\times 10^{46}$~s\e.  The mass at
which the ionizing photon luminosity exceeds this value depends on
$\fkep$; for the fiducial case of $\fkep=0.5$, this occurs at about
$m_*\simeq 20\sm$. Second, we have assumed that in the outer parts of
the \ion{H}{2} region, where the helium is singly ionized for stellar
temperatures $\sim 10^5$~K, each recombination of He$^+$ results in
one H ionization, whereas it actually results in about 2/3 of an
ionization at the relevant densities ($>10^4$ cm\eee; Osterbrock
1989).  In fact, the abundance of He is sufficiently small ($\sim
0.08$) that we shall henceforth neglect it in our analytic estimates
(however, we do not neglect its contribution to the mass density, nor
is it neglected in the numerical calculations).
Finally, we have ignored photoionization from the
$n=2$ level of H, so that our calculation somewhat
underestimates the size of the ionized region, although this is not very
important at the densities resulting from realistic values
of $f_{\rm Kep}$.

        With the density distribution given by equation (\ref{eq:rho}),
equation (\ref{eq:somega}) becomes
\beq
S=\frac{\atwo \mdsd^2 I}{8\pi \muh^2G \msd}\equiv \scr I,
\label{eq:si}
\eeq
where $\muh=2.20\times 10^{-24}$~g is the mass per hydrogen
and
\beq
I\equiv \int_{r_*}^{\rtwo}\frac{\psi^2(\mu,r)}{r}\ dr.
\eeq
We have set $m=\msd$ in equation (\ref{eq:rho}) 
in accord with the discussion following equation
(\ref{eq:rdr}).
Equation (\ref{eq:si}) reduces to that of Omukai 
\& Inutsuka (2002) for $\psi=\ln(\rtwo/r_*)$ (and if $\muh$ is
replaced by $m_p$, $\mdsd$ by $\mds$ and $\msd$ by $m_*$).
As shown by Omukai \& Inutsuka, an \ion{H}{2} region in an
$r^{-3/2}$ density profile 
is confined to the vicinity of the star
for $S\la \scr$ and expands to exponentially large
distances as $S$ increases above $\scr$.
Numerically, we have
\beq
\scr=3.07\times 10^{50}\left(\frac{2.5}{T_4}
        \right)^{0.8}\;\frac{\mdsdt^2}{\msdt}
        ~~~~{\rm ph\ s^{-1}} \rightarrow
        2.36\times 10^{51}\left(\frac{2.5}{T_4}
        \right)^{0.8}\;\frac{K'^{30/7}}{\mst^{13/7}}
        ~~~~{\rm ph\ s^{-1}}.
\label{eq:scr}
\eeq
By comparison, the ionizing luminosity of a Pop III star is
\beq
S\simeq 7.9\times 10^{49}\; \phi_S \mst^{1.5}~~~~{\rm ph\ s^{-1}},
\label{eq:ss}
\eeq which for $\phi_S=1$ is a fit to Schaerer's (2002) results for
the ionizing luminosity of main sequence primordial stars; the fit is
accurate to within about 5\% for $60 M_\odot \la m_*\la 300 M_\odot$.
As shown in Paper I, the ionizing luminosity can be less than the main
sequence value ($\phi_S<1$) when the star is still contracting toward
the main sequence, and it can be greater when it is accreting while on
the main sequence; for the case illustrated in Paper I, $\phi_S\la 2$.
If the accretion rate is not reduced by feedback effects, $S$ would
not exceed $\scr$ until $m_*>275 K'^{14/47}M_\odot$ for $T_4=2.5$.
However, as we shall see, rotation makes the factor $I$ small and
allows the \ion{H}{2} region to expand until it is almost as large as the
disk even when the mass is less than this.

        At early times we have $\rtwo\ll r_d$ so that we can use
approximation (\ref{eq:psiaa}) for the density. As a result,
we find
\beq
\frac{S}{\scr}=I\simeq \frac{1}{4(1+\mu)}\left(\frac{\rtwo}{r_d}\right)^2.
\eeq
With equations (\ref{eq:scr}) and (\ref{eq:ss}), we
then obtain
\beqa
\frac{\rtwo}{r_d} & = & 1.01(1+\mu)^{1/2}(1+f_d)^{1/2}
	\phi_S^{1/2}
        \left(\frac{T_4}{2.5}\right)^{0.4}\;\frac{\mst^{1.25}} 
        {\mdsdt},       \\
        & \rightarrow & 0.37\; (1+\mu)^{1/2}
	\phi_S^{1/2}
        \left(\frac{T_4}{2.5}\right)^{0.4}\;\frac{\mst^{47/28}}
        {K'^{15/7}}\; ;
\label{eq:rtwod}
\eeqa
recall that ``$\rightarrow$'' indicates that
we have taken $\msd=\frac 43 m_*$.
We see that for $\mst\la 1$ we have $\rtwo\la 0.5 r_d$, so
that our approximation for the density, equation (\ref{eq:psiaa}), is
reasonably accurate at early times. The radius of the \ion{H}{2} region is then
\beq
\rtwo=5.40\times 10^{15}\; (1+\mu)^{1/2}
	\phi_S^{1/2}
        \left(\frac{T_4}{2.5}\right)^{0.4}\left(\frac{\fkep}{0.5}\right)^2
        \frac{(1+f_d)^{31/14}\mst^3}{K'^{25/7}}~~~{\rm
cm}~~~~~~(\mst\la 1),
\label{eq:rtwo}
\eeq where we have set the star formation efficiencies $\esd$ and
$\esdb$ equal to unity and where we have made the approximation
$83/28\simeq 3$ in the exponent of $\mst$. As the stellar mass
increases above $100 M_\odot$, the approximation for the density,
equation (\ref{eq:psiaa}), increasingly overestimates the density
except near the equator; as a result the radius of the \ion{H}{2} region,
$\rtwo$, becomes larger than the value given in equation
(\ref{eq:rtwo}) except near the plane of the disk, where the high
density traps the \ion{H}{2} region.  As remarked above, for $S>\scr$, which
occurs for $\mst>2.75$ if the accretion continues unabated by
feedback, $\rtwo$ increases exponentially with $S$ (Omukai \& Inutsuka
2002).

\subsubsection{Later Evolution of the \ion{H}{2} Region: Suppression
of Accretion}
\label{S:later}

        According to equation (\ref{eq:rtwo}), the radius
of the \ion{H}{2} region expands on the protostellar evolution timescale
$\sim 10^4$ yr, which is far longer than the dynamical time
$\rtwo/2c_i\sim 10^2$~yr. As a result, the velocity of the ionization
front relative to the infalling
gas is very nearly equal to the free-fall velocity.
The first phase of evolution of the \ion{H}{2} region ends when
it expands to the point that 
the radius becomes comparable to the gravitational radius,
\beq
\label{eq:rg}
r_g  \equiv \frac{G\phiedd\msd}{c_i^2}=
        3.92\times 10^{15}\phiedd\left(\frac{2.5}{T_4}\right)\msdt
        ~~~{\rm cm},
\eeq
where we have taken the gravitating mass to be $\msd$.
Here we have allowed for the decrease in the radiation pressure
due to electron scattering through the factor
\beq
\phiedd\equiv 1-\frac{L}{L_{\rm Edd}}\; ,
\label{eq:phiedd}
\eeq
where $\ledd=
4\pi G m c/\kappa_{\rm Thomson}$ is the Eddington limit. 
In Paper I we found that $L/\ledd\sim 0.6-0.8$ 
for $m=m_*\sim 10^2\, \sm$, which corresponds to $\phiedd\simeq 0.2-0.4$.
Equations (\ref{eq:rtwo}) and (\ref{eq:rg}) relate the protostellar
mass to $\rtwo/r_g$,
\beq
\mst  =  0.85\left[\frac{\phiedd^{1/2}K'^{25/14}}{(1+\mu)^{1/4}(1+f_d)^{17/28}
\phi_S^{1/4}
}\right]
        \left(\frac{2.5}{T_4}\right)^{0.7}
        \left(\frac{0.5}{\fkep}\right)\left(\frac{\rtwo}{r_g}\right)^{1/2}\; .
\label{eq:rfront}
\eeq
Keep in mind that this relation is valid only for $\mst\la 1$, so that $\fkep$ cannot
be small. This condition is  satisfied
insofar as the simulations of ABN are representative
of the angular momentum of the accreting gas,
since they give $\fkep\sim 0.5$.

	When the \ion{H}{2} region expands to the point that $\vff=
2c_i$, a shock front forms and the ionization front becomes D-type;
this occurs at $\rtwo=r_g/2$. The accretion rate through the
\ion{H}{2} region will begin to decrease at this point.  Since the
shocked neutral gas is denser than the ionized gas in the \ion{H}{2} region,
the accelerating expansion of the shocked shell is subject to the
Rayleigh-Taylor instability, and as a result it is difficult to
estimate by how much the accretion is reduced.  While the shell is
moving slowly, it can fall onto the disk and accrete that way.
However, once the shocked shell is moving faster than the local
free-fall velocity it seems unlikely than any significant further
accretion can occur.  To obtain an approximate upper limit on the
accretion through the \ion{H}{2} region, we assume that, from a given
direction, the accretion is unimpeded until the \ion{H}{2} region has
expanded to a radius equal to $r_g$.  Because of the declining density
distribution in the infall envelope, the \ion{H}{2} region is
expanding relatively rapidly at this stage and so soon ionizes a large
region beyond $r_g$, which we expect leads to a substantial reduction
in accretion rate from the affected directions. This approximation
needs to be checked with multi-dimensional radiation-hydrodynamical
simulations.  It is important to bear in mind that this suppression of
the accretion occurs only in the ionized gas. Gas in the shadow of the
accretion disk around the star can continue to accrete, as discussed
below.

\begin{figure}[h]
\begin{center}
\epsfig{
        file=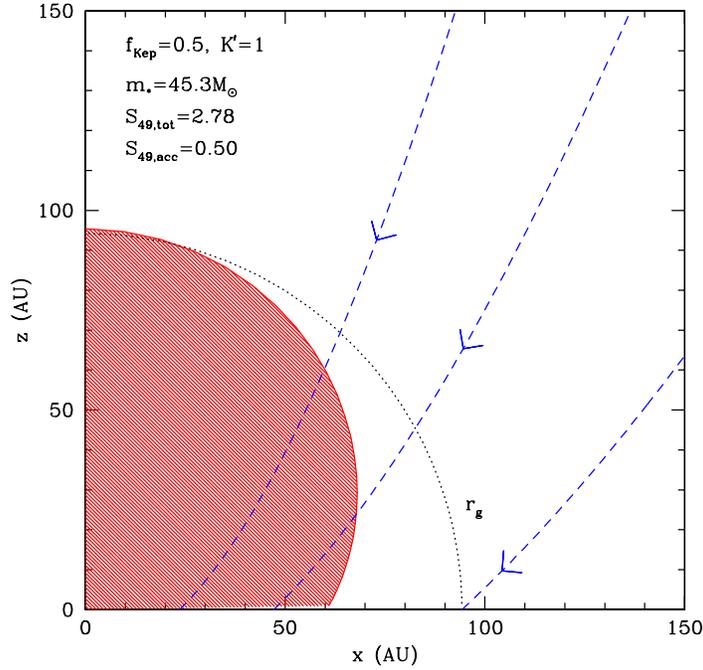,
        angle=0,
        width=\figwidth}
\end{center}
\caption{
\label{fig:hiiplotmary2}
Geometry of the \ion{H}{2} region (shaded) assuming the sector
approximation (see text) during the breakout phase for the fiducial
model with $K^\prime =1$ and $f_{\rm Kep}=0.5$. The protostar is at
(0,0) and the disk is in the $z=0$ plane. At this stage, the star has
$m_*=45 M_\odot$, a total ionizing photon luminosity of $S_{49, \rm
tot}=2.78$, of which $S_{49, \rm acc}=0.50$ comes from accretion. The
\ion{H}{2} region has just recently expanded beyond $r_g$ (at 94 AU)
in the polar direction. The geometry of several accretion streamlines
is shown by the dashed lines.
}
\end{figure}

\begin{figure}[h]
\begin{center}
\epsfig{
        file=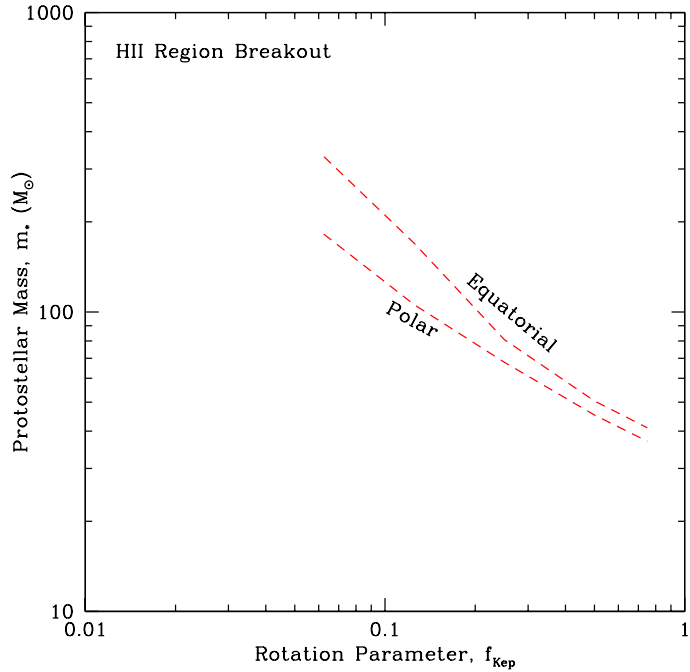,
        angle=0,
        width=\figwidth}
\end{center}
\caption{
  \label{fig:hiibreakout}
Mass scales of \ion{H}{2} region breakout as a function of the
rotation parameter $f_{\rm Kep}$. The lower dashed line marked
``Polar'' shows the mass scale of the protostar at which the
\ion{H}{2} region reaches $r_g$ (based on star plus disk mass) along
the rotation axis of the protostar. The upper dashed line marked
``Equatorial'' shows the mass scale of the protostar when the
\ion{H}{2} region reaches $r_g$ in a direction just above the disk
plane ($0.9\pi/2$ from the rotation axis). Note this condition for
equatorial \ion{H}{2} region breakout ignores the effects of reduced
accretion rates from prior polar \ion{H}{2} region breakout, although
such effects are accounted for in the full feedback+evolution models presented below.}
\end{figure}

In figure \ref{fig:hiiplotmary2} we show the geometry of the
\ion{H}{2} region near the point of polar breakout of the ionized gas
beyond $r_g$.  In this calculation the protostellar evolution has been
followed as described in Paper I, but now including the effects of a
reduction in accretion rate once the \ion{H}{2} region breaks out
beyond $r_g$. This has only just started to occur at the point of the
evolution shown in the figure. We have assumed there is negligible
reduction in the accretion rate because of the \lal\ radiation
feedback since we expect its effects to be limited to relatively small
angles around the polar axis.  The extent of the \ion{H}{2} region is
calculated using the sector approximation using the density
distribution model of Ulrich (1976) as described above.  We include
the effect of electron scattering, but not force due to
photoionization, which is discussed below. The effect of radiation
pressure due to photoionization is strongest for purely radial infall,
so its neglect is not critical for the models presented here. We do
allow for advection of neutrals into the \ion{H}{2} region, though
they are not very important by the time the protostar is $\sim
100\:M_\odot$. A temperature of $2.5\times 10^4\:{\rm K}$ was adopted
for the ionized gas, which affects the value of $r_g$.  Note that in
figure \ref{fig:hiiplotmary2} we have assumed an infinitely thin
accretion disk.  The polar and equatorial breakout conditions are
shown as a function of $f_{\rm Kep}$ in figure \ref{fig:hiibreakout}.
Once the protostar
has reached the masses indicated by the ``Equatorial'' line in this
figure, we do not expect accretion to be able to proceed from
directions that have a direct line of sight to the protostar,
i.e. those directions that are not shielded by the accretion
disk. Thus 
in most cases
we do not expect \ion{H}{2} region breakout
to set the final mass of the star, but rather to cause a decrease in
accretion efficiency that starts to become important at about
$50\:M_\odot$ in the fiducial case. The actual reduction in accretion
efficiency depends on the thickness of the accretion disk, to be
discussed below (\S\ref{S:shadow}).

We can compare the analytic prediction for \ion{H}{2} region breakout
(eq. \ref{eq:rfront}) with our numerical calculation, which for the
fiducial model ($f_{\rm Kep}=0.5$, $K^\prime=1$, $T_4=2.5$, $f_d=1/3$)
finds breakout in the polar direction at a mass of 45.3~$\sm$. At this
point the total H-ionizing luminosity is $S_{49} = 2.78$ so that
$\phi_S=1.15$ and the bolometric luminosity is $5.95\times
10^5\:L_\odot$ so that $\phi_{\rm Edd}=0.59$. With these values, the
analytic estimate for the mass at which polar ($\mu=1$) \ion{H}{2}
region breakout ($r_{\rm HII}=r_g$) occurs (eq. \ref{eq:rfront}) is
44.7~$\sm$, in excellent agreement with the numerical value.  
In Fig.~\ref{fig:hiibreakout} we see that the mass at which the
\ion{H}{2} region breaks out does not scale as a simple power of
$f_{\rm Kep}$. This is because $\phi_{\rm Edd}$ and $\phi_S$ vary with
stellar mass, especially for $m_*\lesssim 100\sm$.

\subsection{Radiation Pressure due to Photoionization}
\label{sec:rppi}

        Continuum radiation is dynamically coupled to
the gas in the \ion{H}{2} region, both through Thomson scattering and through
photoionization. Since the \ion{H}{2} region is optically
thin to Thomson scattering, it effectively reduces the force
of gravity by a factor 
$\phiedd=1-L/\ledd$, which as discussed above is
$\sim 0.2-0.4$ for $\mst\sim 1$.
At distances large enough that the mass acting on
the gas is $\msd$, $L/\ledd$ is reduced by a factor
$1+f_d$. Keep in mind that the decrease in the 
effective gravity due to electron scattering reduces the accretion rate of ionized
gas only; it does not affect the accretion of neutral gas
outside the \ion{H}{2} region onto the disk.

        Every photoionization results in a transfer of
momentum $h\nu_i/c$ to the gas, where $h\nu_i$ is
the mean energy of the photons that ionize the gas.
The importance of radiation pressure associated with
photoionization has long been appreciated in
studies of active galactic nuclei and
X-ray sources (e.g., Tarter
\& McKee 1973); Haehnelt (1995) pointed out its
importance in the formation of the first galaxies, and
Omukai \& Inutsuka (2002) discussed its role in
the formation of the first stars.  
They showed that, under the assumption of
perfect spherical symmetry, radiation pressure
would have the counter-intuitive effect of
{\it reducing} the size of the \ion{H}{2} region,
thereby eliminating any feedback effect on
the growth of the protostar. 
Since photoionizations are balanced by recombinations,
the radiative force is given by $\atwo n_p^2(h\nu_i/c)$.
Generalizing their treatment to include electron scattering, we find that
this radiative force
balances the effective gravity at a critical density $\ncr$
given by
\beq
\atwo \ncr^2 \left(\frac{h\nu_i}{c}\right)=\frac{\phiedd\rho_{\rm cr} G\msd}{r^2},
\eeq
so that
\beq
\ncr=2.15\times 10^6 \phiedd\left(\frac{T_4}{2.5}\right)^{0.8}
        \frac{\msdt}{r_{16}^2}~~~~{\rm cm}^{-3},
\eeq
where we have assumed a typical ionizing photon energy
of $1.5\times 13.6 $~eV.
For gas accreting in free-fall 
(i.e., it enters the \ion{H}{2} region at a velocity
that is unaffected by radiation pressure), 
this corresponds to a critical radius
\beqa
r_{\rm cr} & = & 2.36\times 10^{14}\phiedd^2 \left(\frac{T_4}{2.5}\right)^{1.6}
        \frac{\msdt^3}{\mdsdt^2}~~~~{\rm cm},\\
        & \rightarrow & 5.49\times 10^{13}\phiedd^2
        \left(\frac{T_4}{2.5}\right)^{1.6}\frac{\mst^{27/7}}{K'^{30/7}}
        ~~~~{\rm cm}.
        \label{eq:rcr1}
\eeqa
Even for $\phiedd=1$, this radius is typically a few AU in size, and
the infall velocity is highly supersonic relative to the sound speed of the ionized
gas.
Omukai \& Inutsuka showed that as the ionizing flux from the
protostar increased, the radius of the \ion{H}{2} region would increase
until it approached $\rcr$. As it did so, the inflow velocity
would drop, the density would rise, and the accreting gas
would be able to absorb a larger number of ionizing photons.
In fact, the accretion flow inside $\rcr$ could absorb
more ionizing photons than any star could emit. As a result,
the \ion{H}{2} region would remain trapped at small radii and the
gas would continue to accrete supersonically onto the star.
It is not clear that this flow would be stable in three dimensions,
however, since a fluctuation that placed an element of ionized gas
at $r>\rcr$ would result in a net outward force on the gas.

        Angular momentum in the accreting gas changes
this picture completely: The density of the infalling
gas is reduced inside the disk radius $r_d$ (eq. \ref{eq:rd2}), which
is generally much larger than $\rcr$:
\beq
\frac{r_d}{\rcr}=509 \left(\frac{K'^{20/7}}{\phiedd^2}\right)\left(\frac{T_4}{2.5}\right)^{-1.6}
        \left(\frac{\fkep}{0.5}\right)^2
        \mst^{-18/7}.
\eeq
For typical masses $m_*\sim 10^2 \sm$, the disk radius
is larger than $\rcr$ only for very low rotation, $\fkep\la 0.02\phiedd$.
 Repeating the analysis that led to equation (\ref{eq:rcr1}) with
 the density appropriate for rotating infall (eqs. \ref{eq:rho}
and \ref{eq:psiaa}), we find
\beqa
\rcr &\simeq& 2^{2/3}\left(\frac{1+\mu}{2}\right)^{1/3}r_{\rm cr,\, spherical}^{1/3}r_d^{2/3}\; , \\
&\simeq&5.5\times 10^{15}
\phiedd^{2/3}\left(\frac{1+\mu}{2}\right)^{1/3}\left(\frac{\fkep}{0.5}\right)^{4/3}
        \left(\frac{T_4}{2.5}\right)^{0.53}\frac{\mst^{15/7}}{K'^{50/21}}
        ~~~~{\rm cm},
\eeqa
where $r_{\rm cr,\,spherical}$ is the critical radius for spherical infall (eq. \ref{eq:rcr1}).
This result shows that in a rotating infall, the critical radius beyond which radiation
pressure dominates the effective gravity is intermediate between the critical radius in the spherical
case and the disk radius. Comparison with equation (\ref{eq:rg}) shows that
in the rotating case, the critical radius is comparable to the gravitational radius,
where pressure effects can drive the outward expansion of the \ion{H}{2} region. 
It therefore appears  that for typical values of the rotation,
radiation pressure due to photoionization cannot
result in the confinement of the \ion{H}{2} region; correspondingly, feedback
by the \ion{H}{2} region cannot be curtailed by this effect.

\section{Disk Shadowing}
\label{S:shadow}

An optically thick accretion disk is able to shield part of the outer
accretion flow from direct protostellar feedback.  In order to
determine how effective this shielding is, we must know the thickness
of the accretion disk.  With a few significant exceptions (e.g.,
Paczynski \& Bisnovati-Kogan 1981; Meyer \& Meyer-Hofmeister 1982;
D'Alessio et al. 1998), almost all the work on accretion disks has
gone into determining their radial structure; the vertical structure
is generally integrated over. Here we shall estimate the thickness of
the disk under the assumption that it is geometrically thin but very
optically thick.  We neglect convective and turbulent transport in the
disk, which D'Alessio et al. found to be small for the cases they
considered.  We focus on the inner parts of the disk, where
self-gravity is unimportant (Tan \& Blackman 2004).  For simplicity,
we neglect heating of the disk by irradiation from the central source;
our estimate of the disk thickness is thus a lower limit to the true
thickness.
 
           The radial structure of the disk is governed
by the equations 
of energy conservation and of angular momentum conservation.
Energy conservation gives the emergent flux as (Paper I)
\beqa
F_0 &=& \frac{\mds}{4\pi\varpi}\frac{\partial}{\partial\varpi}
        \left(\frac 53 \bar\epsilon_{\rm th}+\bar\epsilon_I\right)
        +\frac{3Gm_*\mds f}{8\pi\varpi^3},\\
    & \equiv & \phi_I\left( \frac{3Gm_*\mds f}{8\pi\varpi^3}\right).
\label{eq:en}
\eeqa
Here 
\beq
f\equiv 1-\left(\frac{\varpi_0}{\varpi}\right)^{1/2},
\label{eq:f}
\eeq
is the factor that embodies the boundary condition that
angular momentum cannot be transferred across a surface
on which the angular velocity has no gradient;
$\varpi_0$ is the cylindrical
radius at which $\partial\Omega/\partial\varpi$
vanishes,
which we take to be equal to the stellar radius.
The dimensionless
factor $\phi_I$  describes the advection of thermal and internal
energy in the disk and is generally less than unity.

       To evaluate the angular momentum transfer
in the disk, we adopt the $\alpha$-disk model of Shakura \& Sunyaev
(1973), in which the transverse stress in the disk is
proportional to the pressure, $w_{\varpi\phi}=-\frac 32\alpha P$
(we have included the factor $\frac 32$ to conform with
convention---Frank et al. 1995).
The equation describing angular momentum transport is then
\beq
\mds\Omega f=6\pi\alpha\int_0^{z_s} Pdz,
\label{eq:angmom}
\eeq
where $z_s$ is the height of the surface of the disk.

          The vertical structure of the disk is governed by three equations:
First is the first moment of the radiative transfer equation,
\beq
\pbyp{\prad}{z}=-\frac{\rho\kappa_F F}{c},
\label{eq:firstmom}
\eeq
where $\kappa_F$ is the flux-weighted mean opacity per unit mass
and $F(z)$ is
the radiative flux. We assume that the effective optical depth
for true absorption, $\tau^*=(\tau_{\rm abs}\tau_{\rm scatt})^{1/2}$,
is much greater than unity so that the gas is approximately in
LTE (Shakura \& Sunyaev 1973; Artemova et al. 1996). 
Then $\prad\simeq\frac 13 aT^4$ and
$\kappa_F\simeq \kapr$, where
$\kapr$ is the Rosseland mean opacity per unit mass,
so that equation (\ref{eq:firstmom})
reduces to the equation of radiative diffusion,
\beq
\pbyp{T}{z} = -\frac{3 \kapr \rho F}{16 \sigma T^3}.
\label{eq:dtdz}
\eeq

    The second equation
describes the growth of the flux due to viscous dissipation,
\beq
\pbyp{F}{z}=-\phi_I w_{\varpi\phi}\varpi\;\pbyp{\Omega}{\varpi}
        =\frac 94 \phi_I\alpha\Omega P
\label{eq:fz}
\eeq
(Shu 1992). We have included the factor $\phi_I$ to
allow for the reduction in the flux
by the advection of internal energy.
In addition to the factor $\frac 32\phi_I$,
equation (\ref{eq:fz}) differs from the expression adopted by
Shakura \& Sunyaev (1973) in that it has $\partial F/\partial z\propto
P$ rather than  $\propto\rho$. One can show, however, that
the height of the disk is very insensitive to this change.
Integration of equation (\ref{eq:fz}) together with equation
(\ref{eq:angmom}) leads directly to the energy equation (\ref{eq:en}).

          Finally, we have the equation of 
hydrostatic equilibrium,
\beq
\pbyp{P}{z}
        =-\frac{\rho Gm_*z}{\varpi^3},
\eeq
where the pressure $P$ includes both gas pressure and radiation pressure,
\beq
P=P_g+\prad=\frac{\rho kT}{\mu}+\frac{4\sigma T^4}{3c}.
\eeq
For a primordial gas with a helium fraction of $0.079$,
the mean mass per particle is
\beq
\mu\simeq \frac{1.32m_{\rm H}}{1.08+x_e}=\frac{2.20\times 10^{-24}}{1.08
        +x_e}~~~~{\rm g},
\eeq
where $x_e\equiv n_e/n_{\rm H}$ is the ionization fraction relative
to hydrogen; for a fully ionized primordial gas, $\mu=0.98\times
10^{-24}$ g.
With the aid of equation (\ref{eq:dtdz}), the equation of
hydrostatic equilibrium becomes
\beq
\pbyp{\rho}{z}=\left(\frac{kT}{\mu}\right)^{-1} 
\left[-\frac{Gm_*\rho z}{\varpi^3} + 
\frac{\rho\kapr F}{c}\left(1+\frac{3kc\rho}{16\mu\sigma T^3}\right)\right].
\label{eq:drhodz}
 \eeq

 This is a two-point boundary value problem, in which
$F=F_0$ at the surface of the disk and $F=0$ at the midplane.
We assume that the disk is very opaque, so that 
we can neglect the flux generated above the photosphere;
we can therefore apply the surface boundary conditions
at the photosphere, located at $\zph$.
Since we have assumed that the disk is opaque and are neglecting
irradiation, the surface temperature is small compared
to the central temperature; as a result, the scale height
near the surface is small and $\zph\simeq z_s$.
The temperature at the photosphere is the effective temperature,
which is related to the emergent flux by $F_0=\sigma T_{\rm eff}^4$.

We have addressed this problem both analytically and numerically.
The case of pure radiation pressure is trivial to treat analytically, since hydrostatic
equilibrium gives
\beq
\frac{\kapr F}{c} = \frac{Gm_*z}{\varpi^3}
\eeq
which is true throughout the disk. At the surface of the disk
(which is denoted $z_{sr}$ for a radiation-pressure supported disk) this gives 
\beq
z_{sr}=\frac{\krs F_0 \varpi^3}{Gm_*c},
\label{eq:zsr}
\eeq
where $\krs$ is the opacity at the disk surface.
With the aid of equation (\ref{eq:en}), this becomes
\beqa
z_{sr} &=& \left(\frac{3\phi_I\krs}{8\pi c}\right)\mds f\\
     &=& 8.77\times 10^{10}\;\left(\frac{\krs}{\kht}\right)\phi_I
     \mdst f~~~~~{\rm cm},
\eeqa
where we assume that $\mds=\mdsd/(1+f_d)\rightarrow\frac 34 \mdsd$
and where $\kht=(n_e/\rho)
\sigma_{\rm T}=0.35$~g~cm\ee\  
is the opacity due to electron scattering 
for fully ionized primordial gas. Note that the thickness 
of a radiation-supported disk depends
only weakly on radius through the factor $f=1-(\varpi_0/\varpi)^{1/2}$
and possibly through a variation in the opacity at the surface.

The case of a disk supported by gas pressure is more complicated,
and is discussed in Appendix E for the case of constant opacity. 
There we show that the height of such
a disk is
\beq
z_{sg}\simeq 1.21\times 10^{10}\left(\frac{\phi_I\bar\kapr}{\alpha_{-2}\kappa_T}
                \right)^{1/10}\left(\frac{\varpi}{R_\odot}\right)^{21/20}
                \frac{(\mdst f)^{1/5}}{\mst^{7/20}}~~~~~{\rm cm},
\eeq
where we have normalized $\alpha$ to a typical value of 0.01.
In the above expression, we have replaced the constant opacity
in the derivation in the Appendix with a suitable mean value.
Observe that the
height of a disk supported by gas pressure scales almost linearly with
$\varpi$, so that the aspect ratio is approximately constant.

        Numerical solution of the structure equations shows
that when both radiation pressure and 
gas pressure are important, the approximation
\beq
z_s\simeq \left(\zsr^{5/4}+z_{sg}^{5/4}\right)^{4/5}
\label{eq:zs}
\eeq
is accurate to better than 10\% over the range
$10^{-4}\la\alpha\la 10^{-2},\ \kapr\simeq\kappa_T,\ 10^4$~K$\la\teff\la
10^6$~K, and $0.01\la\zsr/z_{sg}\la 10$, provided the opacity is constant.

\begin{figure}[h]
\begin{center}
\epsfig{
        file=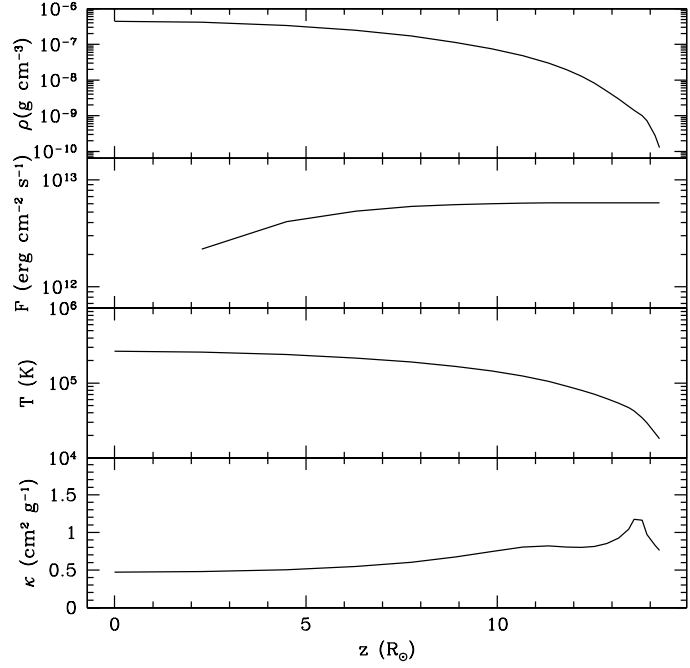,
        angle=0,
        width=\figwidth}
\end{center}
\caption{
\label{fig:struct}
Vertical structure of the accretion disk at $r=10r_*\simeq 43R_\odot$ for
$m_*=100\:M_\odot$, $K^\prime=1$, $f_{\rm Kep}=0.5$ and no reduction
in accretion efficiency due to feedback.  }
\end{figure}

	We have also solved the equations for the vertical structure of the disk
numerically with a realistic opacity
variation with density and temperature (Iglesias \& Rogers 1996).
We follow the disk structure 
during the course of the protostellar evolution (i.e. as
$m_*$, $\dot{m}_*$ and $r_*$ evolve).  We
adopt an $\alpha$-viscosity parameter of 0.01, 
typical of values associated with the magnetorotational instability (Balbus \& Hawley 1998; Tan \& Blackman 2004),
although the
disk thickness is not very sensitive to this choice. An example of the
vertical disk structure is shown in figure \ref{fig:struct} for a
location at 10$R_*$ around a 100$\sm$ main sequence star, accreting at
$2.4\times 10^{-3}\:M_\odot\:{\rm yr^{-1}}$, i.e. the fiducial rate
from a $K'=1$ core with no reduction due to feedback. The numerical
value of $z_s/\varpi$ is 0.33. This compares with an analytic
estimate based on equation (\ref{eq:zs}) of 0.31, where we adopted
$\krs=0.75\:{\rm cm^2\:g^{-1}}$ 
and $\bar\kappa_{\rm R}=0.6\:{\rm cm^2\:g^{-1}}$ 
(Fig. \ref{fig:struct}).

\begin{figure}[h]
\begin{center}
\epsfig{
        file=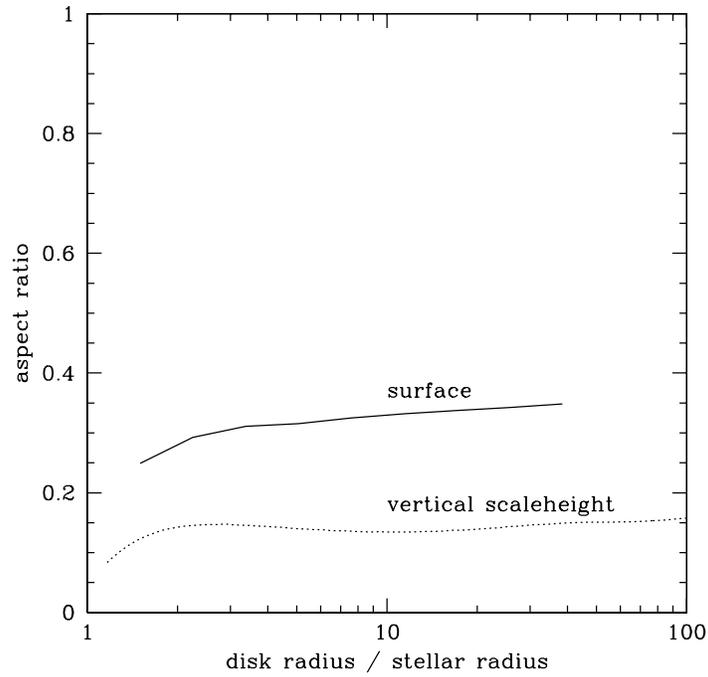,
        angle=0,
        width=\figwidth}
\end{center}
\caption{
\label{fig:radial}
Radial dependence of the aspect ratio of the disk surface, $z_s/\varpi$, and disk
density vertical scaleheight, $h/\varpi$, for $m_*=100\:M_\odot$, $K^\prime=1$, $f_{\rm
Kep}=0.5$ and no reduction in accretion efficiency due to feedback.  }
\end{figure}

The radial variation in the density scaleheight and disk surface
height for the above model is shown in figure \ref{fig:radial}. Note
that $z_s/\varpi$ is approximately constant with $\varpi$. For
simplicity, in our numerical models we evaluate $z_s/\varpi$ at a
radial location $\varpi=10 R_*$ and use this to evaluate the fraction
of the sky shielded by the disk. In the example shown in figures
(\ref{fig:struct}) and (\ref{fig:radial}), $z_s/\varpi = 0.33$ at this
location, and the fraction of the sky shadowed by this disk
photosphere is $f_{\rm sh}=0.31$. If the matter at infinity were
spherically distributed in the envelope, then this would be the
approximate accretion efficiency once the \ion{H}{2} region had
expanded to large distances (and assuming all material in the shadow
of the disk remained neutral).  Note that the disk model in the
numerical example given above is somewhat thicker than would be present
around a 100 solar mass star accreting from a $K^\prime=1$ core, since
the accretion rate would have been reduced by feedback. Our numerical
models of the growth of the protostar account for such effects
self-consistently.


\section{Disk Photoevaporation}\label{S:diskevap}

As we have seen, in the presence of rotation the various feedback
mechanisms discussed above will first disrupt infall in the polar
direction and may leave behind much of the material in the equatorial
plane. Material close to the plane will be shielded from the feedback
effects by the formation of an accretion disk.  However, this gas is
subject to disk photoevaporation, and accretion will cease when the
photoevaporation rate exceeds the accretion rate onto the star-disk
system.

To estimate when this criterion is reached, we apply the model of
Hollenbach et al. (1994) to estimate the rate of photoevaporation from
the disk. This rate is calculated
assuming a steady disk with no infall from above or below. The diffuse
ionizing flux, reprocessed through the flared atmosphere of the disk,
illuminates and ionizes material near and beyond $r_g$. 
Hollenbach et al. considered the possibility that the
disk would be flattened due to a stellar wind, but we
shall use the results of their weak wind case. 
The photoevaporation rate
is calculated via
\beq
\label{eq:evap1}
\dot{m}_{\rm evap}  =  2 \muh v \int_{r_g}^{\infty} 2\pi n_0(r) r
\:dr,
\eeq
where the flow velocity, $v$, is set equal to the ionized gas sound speed,
$c_i=18.4(T_4/2.5)^{1/2}\;\kms$, and $n_0$ is the density of ionized
gas at the base of
the ionized disk atmosphere. 
Their analysis gives
\beq
\dot{m}_{\rm evap} 
=4.1\times 10^{-5} S_{\rm 49}^{1/2}T_4^{0.4}\msdt^{1/2}~~\smyr.
\eeq
As they acknowledge, this result is quite approximate, and a
numerical study of this problem would be worthwhile.
For primordial stars with an ionizing luminosity given by equation
(\ref{eq:ss}), the photoevaporation rate becomes
\beq
\dot{m}_{\rm evap} = 1.70\times 10^{-4}\phi_S^{1/2}(1+f_d)^{1/2}\left(\frac{T_4}{2.5}\right)^{0.4}
        \mst^{5/4}~~\smyr.
\label{eq:evap2}
\eeq

        There are two corrections that could be applied to this result.
First, Begelman, McKee, \& Shields (1983) and Woods
et al. (1996) showed that for analogous winds from AGN disks, the flow
can start from radii well inside $r_g$. Numerically integrating
the expression given by Woods et al., we find that mass loss
inside $r_g$ increases the total mass loss by a factor 1.5.
On the other hand, radiation pressure due to electron
scattering reduces the effective mass
by a factor $\phiedd\sim 0.3$ (eq. \ref{eq:phiedd}).
Since the mass loss rate scales as the
square root of the gravitational mass, this reduction approximately
cancels the increase due to mass loss from the inner disk.
We therefore adopt the Hollenbach et al. estimate of
$\dot m_{\rm evap}$ 
for our analytic 
and numerical 
estimates.

	Accretion onto the star will cease shortly after the photoevaporation
rate exceeds the accretion rate onto the star-disk system, which
is given by equation (\ref{eq:mdsd}). From equation (\ref{eq:evap2}), we
find that  the resulting maximum stellar mass is
\beq
\label{eq:maxevap}
{\rm Max}\; m_{*f,2}=
6.3\;\frac{\esd^{28/47}\esdb^{12/47}
K'^{60/47}}{\phi_S^{14/47}(1+f_d)^{26/47}} 
        \left(\frac{2.5}{T_4}\right)^{0.24}\; .
\eeq
Recall that $\esd$ is the instantaneous star formation efficiency---i.e., the ratio
of the accretion rate onto the star to the rate that would have occurred in the absence of feedback. In the present case, this ratio is just the shadowing factor,
$\fsh$, introduced in the last section. In the numerical solution described below,
we keep track of $\esd$ as a function of time; for the analytic case, we make the simple
approximation that the shadowing sets in when the stellar mass reaches $m_1$, so
that $\esd=1$ until the mass of
the central star reaches $m_1$ and $\esd=\fsh$ thereafter. It is then straightforward to
show that
\beq
\esdb=\frac{\fsh}{1-(1-\fsh)(m_1/\msd)},
\eeq
provided that $\msd\geq m_1$. Note that the average efficiency $\esdb=1$
at the onset of shadowing $(\msd=m_1)$ and that $\esdb\rightarrow\fsh$
at late times $\msd\gg m_1$. Normalizing $\fsh$ to a typical value of 0.2
from the results of \S\ref{S:diskevap} and allowing for smaller accretion rates due to feedback, we find
\beq
\label{eq:maxm}
{\rm Max}\; m_{*f,2}=1.45\, K'^{60/47}
        \left(\frac{2.5}{T_4}\right)^{0.24}\left(\frac{\fsh}{0.2}\right)^
        {28/47}\left(\frac{\esdb}{0.25}\right)^{12/47},
\eeq
where we have set the ionizing luminosity factor $\phi_S=1$ and 
the disk mass fraction 
$f_d=\frac 13$; 
we have normalized
$\esdb$ to a value of 0.25, 
which is approximately correct for $K'=1$ and for
$m_1\simeq 50 M_\odot$ as found in \S\ref{S:later} and $\msd=200\sm$. This analytic
estimate therefore suggests that for the fiducial case ($K'=1$) the
mass of a first generation star should be of order $140 M_\odot$.
We now confirm this with more accurate numerical integrations.

\begin{figure}[h]
\begin{center}
\epsfig{
        file=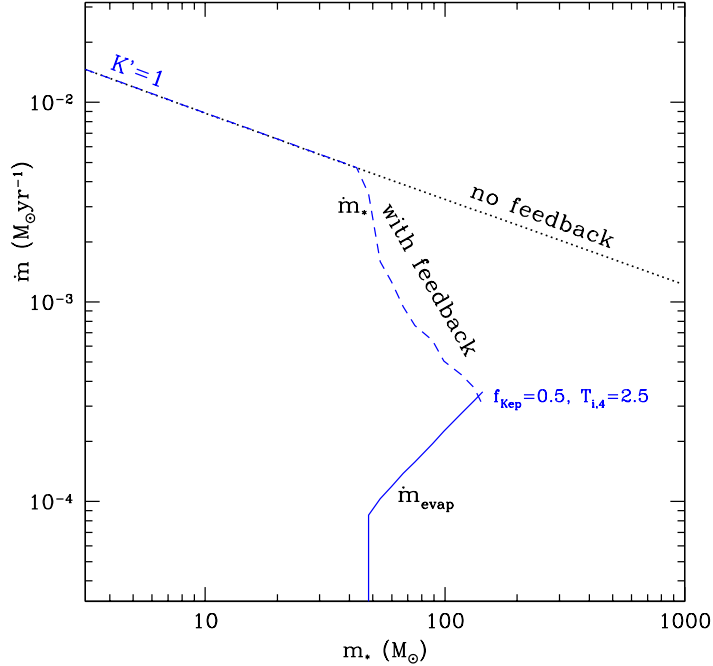,
        angle=0,
        width=\figwidth}
\end{center}
\caption{
\label{fig:theend1}
Feedback-limited accretion: fiducial case. The evolution of the
accretion rate versus protostellar mass is shown for the fiducial
model ($f_{\rm Kep}=0.5$, $K^\prime =1$, $T_i = 25,000$~K) in the
cases of ``no feedback'' and ``with feedback''. In the latter, the
accretion efficiency is reduced as the \ion{H}{2} region expands to
$r_g$ and beyond. However, accretion is allowed to continue from
directions that are shadowed by the disk photosphere. The disk
structure and protostellar structure and feedback are calculated
self-consistently given the evolution in $\dot{m}_*$. Also shown is
the photoevaporative mass loss rate, $\dot{m}_{\rm evap}$, which
starts once the \ion{H}{2} region has broken out in the equatorial
direction and grows as the ionizing flux increases. We see that this
mass loss rate becomes greater than the accretion rate at $m_*\simeq
137\:M_\odot$, and we identify this mass scale as our best estimate of
initial mass scale of the first stars.  }
\end{figure}

We evaluate the photoevaporative mass loss rate in our numerical model
with feedback (Figure \ref{fig:theend1}). The accretion rate is
reduced as the \ion{H}{2} region, with $T=2.5\times 10^4\:{\rm K}$,
expands to $r_g$ and beyond, although accretion is assumed to continue
from directions shielded by the photosphere of the accretion disk. The
disk and protostellar structure and feedback are calculated
self-consistently given this evolution in the accretion rate. Beyond
about 45~$M_\odot$ the accretion efficiency starts to be reduced below
unity. By about 137~$M_\odot$ the evaporative mass loss rate has
become greater than the accretion rate and we then expect very limited
growth of the protostar. We identify this mass scale as our fiducial
estimate for the initial mass of the first stars. At this stage
$\fsh\simeq 0.19$ and $\phi_S=1.37$.\footnote{Note that only about 10\%
of this excess H-ionizing photon production rate is due to
accretion. The remainder is due to our assumption the spectrum of the
protostar can be approximated as a blackbody, rather than the more
detailed stellar atmosphere models of Schaerer (2002).}  This
estimate of the mass at which accretion end agrees well with the
analytic estimate of equation (\ref{eq:maxm}).

\begin{figure}[h]
\begin{center}
\epsfig{
        file=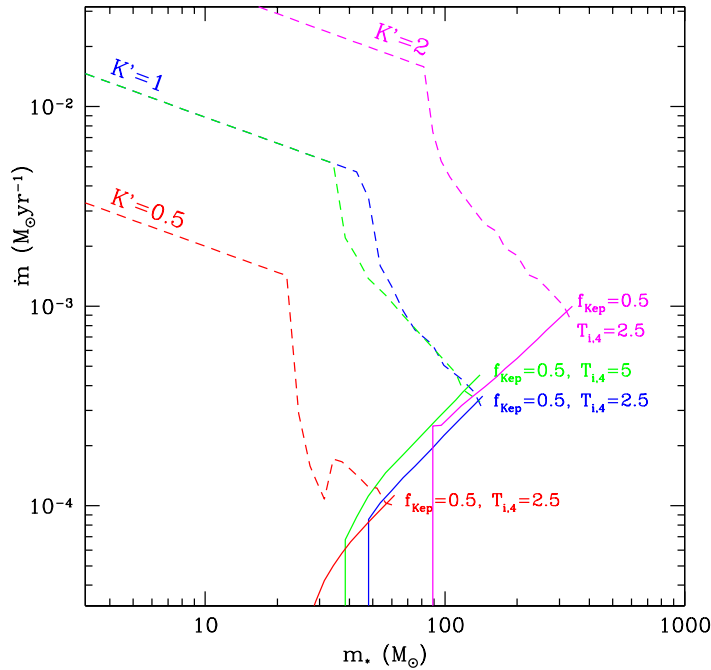,
        angle=0,
        width=\figwidth}
\end{center}
\caption{
\label{fig:theend2}
Feedback-limited accretion: effect of ionized gas temperature and
accretion rate. The fiducial model ($f_{\rm Kep}=0.5$, $K^\prime =1$,
$T_{i,4} = 2.5$~K) shown in Fig.~\ref{fig:theend1} is compared to
models in which one parameter has been changed: a model with
$T_{i,4}=5$ and two models with $K^\prime=0.5,2$. The dashed lines
show the accretion rate to the star, $\dot{m}_*$, and the solid lines
show the photoevaporative mass loss rate, $\dot{m}_{\rm evap}$. The
change in temperature causes relatively minor differences, while the
change in $K^\prime$, equivalent to a change in $\dot{m}_*$ of factors
of 4.4 above and below the fiducial level, leads to roughly a factor
of 2.4 change in the final stellar mass. Note the increase in
$\dot{m}_*$ for the $K^\prime=0.5$ case at around $35\sm$ is due to a
thickening of the inner accretion disk as the star contracts down to
its main sequence configuration and assumes material at large
distances still remains to be accreted in the enlarged shadowed
region.}
\end{figure}

We investigate the sensitivity of this result to the ionized gas
temperature by setting this equal to 50,000~K
(Fig.~\ref{fig:theend2}). This causes the \ion{H}{2} region to break
out sooner and for the disk photoevaporative mass loss rate to be
higher. However, with the other parameters unchanged ($K^\prime=1$,
$f_{\rm Kep}=0.5$), this has only a modest effect on the final mass,
reducing it from 137~$M_\odot$ to 120~$M_\odot$.

We also consider the effect of changing the entropy parameter of the
initial core by factors of two to higher and lower values
($K^\prime=0.5,2$) (Fig.~\ref{fig:theend2}). This corresponds to a
change in accretion rate of factors of 4.4 since $\dot{m}_* \propto
(K^\prime)^{15/7}$. \ion{H}{2} region breakout is accelerated/delayed
by about a factor of two in protostellar mass by these
changes. The final stellar mass set by disk photoevaporation shows a
slightly broader range of factors of 2.4 smaller/greater than the
fiducial value. This is consistent with equation (\ref{eq:maxevap}),
which would predict a change of a factor of 2.4 if the disk thickness
was constant and accretion ionizing luminosity negligible.

\begin{figure}[h]
\begin{center}
\epsfig{
        file=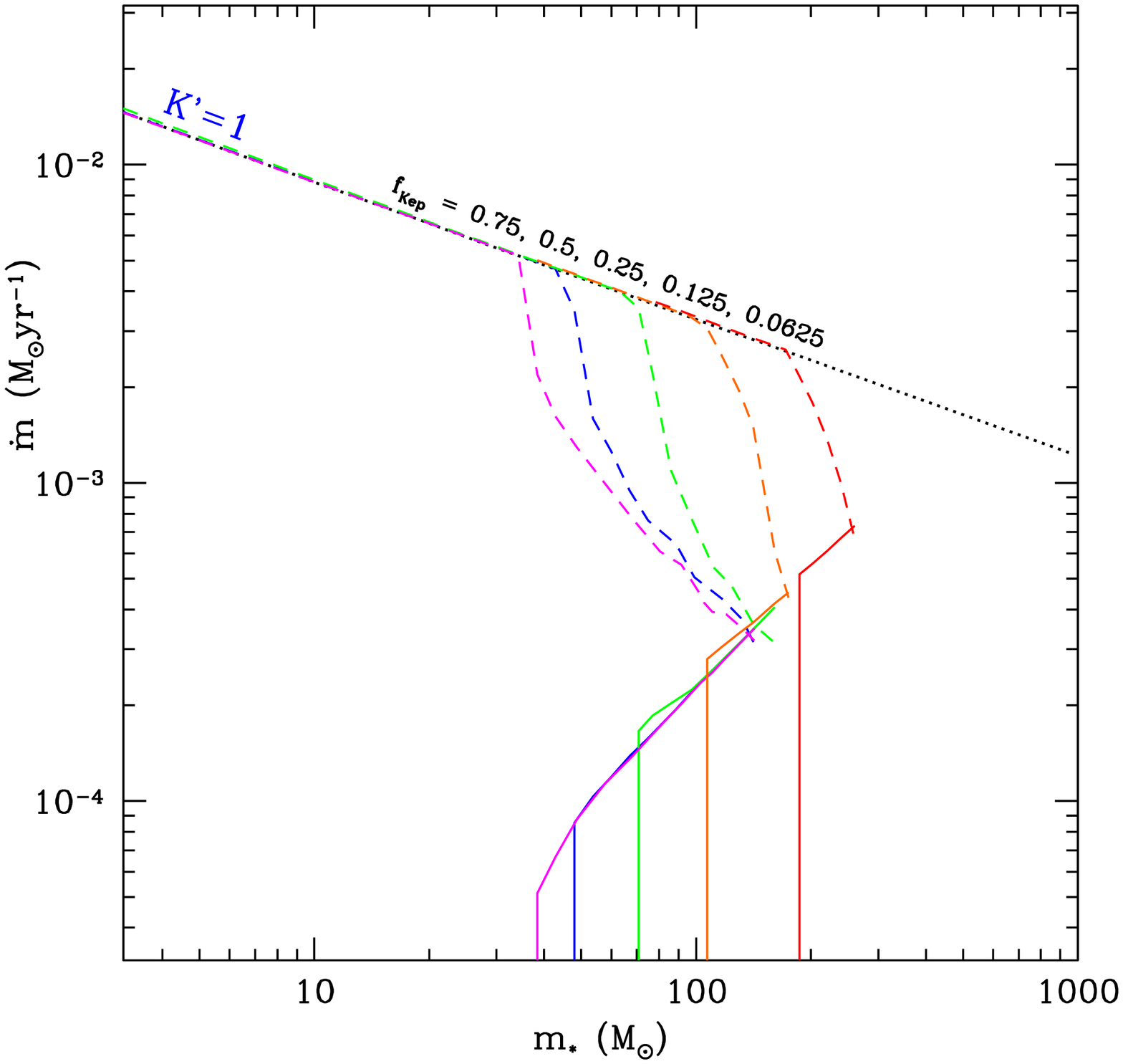,
        angle=0,
        width=\figwidth}
\end{center}
\caption{
\label{fig:theendfKep}
Feedback-limited accretion: effect of rotation. The fiducial model
($f_{\rm Kep}=0.5$, $K^\prime =1$, $T_{i,4} = 2.5$~K) shown in
Fig.~\ref{fig:theend1} is compared to models in which only the
rotation parameter $f_{\rm Kep}$ has been changed: $f_{\rm
Kep}=0.0625, 0.125, 0.25, 0.75$. Smaller rotation parameters result in
higher polar gas densities in the infall envelope and thus delayed
\ion{H}{2} region breakout (Fig.~\ref{fig:hiibreakout}). However for
$f_{\rm Kep}\gtrsim0.25$ this has relatively little effect on the
final mass, which is set by disk photoevaporation (note the
convergence of the $f_{\rm Kep}=0.25, 0.5, 0.75$ models). At smaller
rotation parameters the process of \ion{H}{2} region breakout plays an
important role in setting the mass scale at which the accretion rate
is truncated to be smaller than the photoevaporative mass loss rate.
}
\end{figure}

Finally we explore the effect of changing the core
rotation. Figure~\ref{fig:theendfKep} shows models with $f_{\rm
Kep}=0.0625,0.125,0.25,0.5,0.75$ for $K^\prime=1$ and
$T_i=25,000$~K. Cores with higher rotation rates have lower densities
in the infall envelope near the star so the \ion{H}{2} region can
break out more easily. However, for rotation parameters $f_{\rm
Kep}\gtrsim0.25$ this makes little difference to the final mass, which
is set by the balance between (inner) disk-shadowed accretion and
photoevaporative mass loss. For smaller rotation parameters ($f_{\rm
Kep}\lesssim0.125$) the process of \ion{H}{2} region breakout does play an
important role in setting the mass scale at which the accretion rate
is truncated to be smaller than the photoevaporative mass loss rate.
However, given the results of numerical
simulations of primordial core formation (O'Shea \& Norman 2007), it
appears that these low values of rotation are very rare.

\section{Conclusions}
\label{S:conclusions}

Recent numerical studies have indicated that the initial conditions
for primordial star formation are dense, massive gas cores in
approximate hydrostatic and virial equilibrium. These physical
properties are set mostly by the microphysics of $\rm H_2$ cooling and
not by the initial cosmological density perturbations.

We have described the rate of collapse of these gas cores as a
function of the entropy parameter of the gas, $K$, and the amount of mass
that already collapsed. This accretion rate is very large, so that once
an optically thick protostellar core forms the star grows very quickly.

We have developed a simplified method for modeling protostellar
evolution and applied the appropriate accretion rate for primordial
protostars. The method allows for the treatment of accretion of gas
with angular momentum, so that part of the accretion occurs via a
disk. Using a realistic degree of rotation for the initial gas core we
find that, after the protostar has grown to about a solar mass,
essentially all of the accretion flow is via the disk and conditions
at the protostar are optically thin, in contrast to the spherical
case. This means that the radiation field that the accretion envelope
is exposed to is significantly hotter so that ionization and FUV
radiation feedback can become important.

We considered the impact of the protostellar feedback on the infalling
envelope. Again rotation is important because it modifies the density
distribution in the vicinity of the star. First we discussed the
effects of photodissociation of \htwo, the primary coolant. We showed
that this does not stop accretion if the protostar has already begun
to form, but it can suppress star formation in the vicinity
(c.f. Glover \& Brand 2001).  Next we considered radiation pressure
feedback due to resonant scattering of FUV radiation in the \lal\
damping wings. As a result of the high column densities of neutral gas
around the \ion{H}{2} region, this radiation is trapped and the
pressure amplified by large factors. This radiation pressure becomes
larger than the ram pressure of the infalling gas in the polar
directions for stellar masses of order 20$\sm$. However, once the
infall is reversed at the poles, the \lal\ photons can escape and the
accretion in other directions proceeds unimpeded.  We then considered
the growth of the ionized region.  Once the expansion velocity of this
region exceeds the free-fall velocity, the accretion is halted.  This
typically occurred at about 50 to 100$\sm$, although it took much
larger masses for cases with little angular momentum.  The ionized gas
is confined to the region above the disk, however, so accretion can
continue in the shadow of the disk. Evaluating this, we found that
shadowing permitted accretion to continue at a rate of about 20-30\%
of that in the absence of the \ion{H}{2} region.  Allowing for
photoevaporation of the disk, we found that the final stellar mass is
about $140 M_\odot$ in the fiducial case.

Table \ref{tab:m} summarizes how the mass scales set by protostellar
feedback depend on model parameters.  The final mass of a Pop III.1
star depends fairly sensitively on the entropy parameter of the
accreting gas (i.e., approximately as $(K^\prime)^{1.3}$), which in
turn determines the overall accretion rate to the star$+$disk,
but not very much on core rotation (for $f_{\rm Kep}\gtrsim 0.25$) or
ionized gas temperature ($T_i$). At very low values of core rotation,
\ion{H}{2} region breakout is delayed until high protostellar masses,
at which point the disk photoevaporation rate soon exceeds the
residual disk-shadowed accretion rate. However, these small values of
$f_{\rm Kep}$ are not very likely to occur in nature.

The final masses predicted by our model overlap the range of masses
necessary to produce pair instability supernovae, $140-260\;M_\odot$
(Heger \& Woosley 2002).  Rotation may lower these limits (S. Woosley,
private comm.). The lack of the expected nucleosynthetic signature of
such supernovae in the abundance patterns of very metal poor halo
stars (Tumlinson et al. 2004), could indicate that such massive Pop
III.1 stars were relatively rare or that they tended to enrich regions
not probed by typical halo stars, perhaps the centers of larger
galactic halos.  The conclusion by Scannapieco et al. (2006) that Pop
III star formation should be fairly widespread in regions now probed
by Galactic halo stars, can be reconciled with the abundance pattern
observations if most of this star formation leads to either Pop III.1
stars from relatively low entropy ($K^\prime \lesssim 1$) gas cores or
Pop III.2 stars that also have a mass scale below the pair instability
threshold (see also the study by Greif \& Bromm 2006). Further work is
required to determine the range of pre-stellar core parameters,
primarily $K^\prime$ and $f_{\rm Kep}$, exhibited in cosmological
simulations, in order to predict the frequency of pair instability
supernovae.

\begin{deluxetable}{cccccc} 
\tablecaption{Mass Scales of Population III.1 Protostellar Feedback\label{tab:m}}
\tablewidth{0pt}
\tablehead{
\colhead{$K^\prime$} & \colhead{$f_{\rm Kep}$} & \colhead{$T_i/(10^4\:{\rm K})$} & \colhead{$m_{\rm *,pb}$ ($\sm$)\tablenotemark{a}} & \colhead{$m_{\rm *,eb}$ ($\sm$)\tablenotemark{b}} & \colhead{$m_{\rm *,evap}$ ($\sm$)\tablenotemark{c}}
}
\startdata
1 & 0.5 & 2.5 & 45.3 & 50.4 & 137\tablenotemark{d}\\
\hline
1 & 0.75 & 2.5 & 37 & 41 & 137\\
1 & 0.25 & 2.5 & 68 & 81 & 143\\
1 & 0.125 & 2.5 & 106 & 170 & 173\\
1 & 0.0626 & 2.5 & 182 & 330\tablenotemark{e} & 256\\
\hline
1 & 0.5 & 5.0 & 35 & 38 & 120\\
1 & 0.25 & 5.0 & 53.0 & 61 & 125\\
\hline
0.5 & 0.5 & 2.5 & 23.0 & 24.5 & 57\\
\hline
2.0 & 0.5 & 2.5 & 85 & 87 & 321\\
\enddata
\tablenotetext{a}{\footnotesize Mass scale of \ion{H}{2} region polar breakout.}
\tablenotetext{b}{\footnotesize Mass scale of \ion{H}{2} region near-equatorial breakout.}
\tablenotetext{c}{\footnotesize Mass scale of disk photoevaporation limited accretion.}
\tablenotetext{d}{\footnotesize Fiducial model.}
\tablenotetext{e}{\footnotesize This mass is greater than $m_{\rm *,evap}$ in this case because it is calculated without allowing for a reduction in $\dot{m}_*$ during the evolution due to polar \ion{H}{2} region breakout.}
\end{deluxetable}

One may ask how the feedback mechanisms we have considered relate to
those that operate in contemporary massive star formation. We note
that the maximum mass attained in our fiducial model of PopIII.1 star
formation is very similar to that inferred observationally in local
massive star clusters (e.g. Figer 2005). However, after decades of
study, it remains unclear whether the maximum mass of stars forming
today is set by feedback or instabilities in very massive stars
(Larson \& Starrfield 1971).  Here, we have argued that the maximum
mass of primordial stars is set by feedback. The primary differences
in the feedback processes then and now are:

(1) Dust.  In contemporary star-forming regions, dust destroys \lal\
photons, eliminating them as a significant pressure. On the other
hand, the dust couples the pressure of the UV continuum radiation to
the gas very effectively, and it remains to be determined whether this
limits the final mass of the star; e.g., Yorke \& Sonnhalter (2002)
find that it does, whereas Krumholz et al. (2005a) have not found
evidence that it does.  Dust also affects the evolution of \ion{H}{2}
regions, absorbing a significant fraction of the ionizing photons in
dense \ion{H}{2} regions (Spitzer 1978), thereby reducing the \ion{H}{2} feedback
discussed in \S 5 and the photoevaporation in \S 7.

(2) Magnetic fields. In contemporary protostars, magnetic fields drive
powerful protostellar winds that drive away a significant fraction of
the core out of which the star is forming (Matzner \& McKee 2000). The
cavities created by these winds allow radiation to escape from the
vicinity of the protostar, significantly reducing the radiation
pressure (Krumholz et al. 2005b).  Tan \& Blackman (2004) considered
the influence of such outflows on Pop III.1 cores, concluding the
instantaneous efficiency of accretion could be reduced by a factor of
about 2 from the no-feedback case in an isotropic core by the time the
star reached 100$\sm$. However, these outflows would not confine
ionizing feedback from the star at these masses, so much of the gas
that could be expelled by outflows would have already been disrupted
by \ion{H}{2} region breakout. We conclude that outflows would have a
relatively minor effect on the results presented here, and that the
masses of the first stars are mostly influenced by radiative
feedback. See Tan \& McKee (2008) for further discussion.

(3) Stellar temperatures and luminosities. Primordial stars were
significantly hotter than contemporary stars, resulting in
significantly greater ionizing luminosities. In addition, the
accretion rates of primordial massive stars are much greater, at least
initially, than those of contemporary massive stars (McKee \& Tan
2002; 2003). Future calculations will show whether feedback can be as
effective in setting the maximum mass of contemporary stars as we have
argued that it is for primordial stars.

\acknowledgements We thank T. Abel, F. Adams, V. Bromm, B. Draine,
J. Goodman, D. McLaughlin, G. Meynet, B. O'Shea, J. Ostriker,
S. Phinney, E. Scannapieco, D. Whalen and N. Yoshida for helpful discussions.  During
the course of this work JCT has been supported by a Spitzer-Cotsen
fellowship from Princeton University, by a Zwicky Fellowship from ETH
Z\"urich and by NSF CAREER grant AST-0645412. The research of CFM is
supported by NSF grants AST-0606831 and PHY-0551164 and by the DOE
through grant DE-FC02-06ER41453.

\appendix

\section{Line Profile with Rayleigh Scattering}
\label{app:rayleigh}

        A complication that occurs in our problem is that the
column density can become so large that Rayleigh scattering
is important. From Jackson (1975), we find that in general the
scattering cross section can be expressed as
\def\phin{\phi_{1\nu}}
\beq
\sigma(\nu)=\bar\sigma \dnd\phin
\eeq
where
\begin{eqnarray}
\bar\sigma &\equiv& \frac{1}{\dnd}\int\sigma(\nu)\, d\nu,
\label{eq:sigmabar}\\
\phin &=& \left(\frac{1}{\pi}\right)
        \frac{(4\nu^4/\nu_0^2)(\Gamma/4\pi)}
        {(\nu_0^2-\nu^2)^2+(4\nu^6/\nu_0^4)(\Gamma/4\pi)^2},
\label{eq:phin}
\end{eqnarray}
and $\Gamma$ is the total spontaneous transition rate out of the 
upper and lower levels of the transition.  For the simple case
of a two-level atom (which can be used for Lyman $\alpha$),
$\Gamma=A_{21}$ is the Einstein $A$ coefficient for 
the transition.
Physically, $\phin$ is the line profile for scattering by an
individual atom; the line profile for a gas, $\phi_\nu$, 
is obtained by convolving
$\phin$ with a Maxwellian distribution.

        We now develop an accurate approximation for $\phin$.   
Defining
\beq
f(\nu)\equiv \frac{4\nu^4}{\nu_0^2(\nu_0+\nu)^2},
\label{eq:fnu}
\eeq
which is unity at line center, we have
\begin{eqnarray}
\phin &=& \left(\frac{1}{\pi}\right)\frac{(\Gamma/4\pi)f(\nu)}{
        \Delta\nu^2+(\nu/\nu_0)^2(\Gamma/4\pi)^2f(\nu)}\\
      &\simeq& \left(\frac{1}{\pi}\right)\frac{(\Gamma/4\pi)f(\nu)}{
        \Delta\nu^2+(\Gamma/4\pi)^2}.
\end{eqnarray}
The approximation of dropping the factor $(\nu/\nu_0)^2f(\nu)$ in the denominator
has an error of order
$\Gamma/(4\pi\nu_0)=A_{21}/(4\pi\nu_0)$, which is less than
$10^{-6}$ for Lyman $\alpha$. At low frequencies
($\nu\ll \nu_0$), we have $\phin\propto f(\nu)\propto \nu^4$, 
the standard frequency scaling for Rayleigh scattering.

        In terms of the normalized frequency
shift $x\equiv \Delta\nu/\dnd$, we have $\sigma(x)=\bar\sigma\phi_{1x}$ 
where
\beq
\phi_{1x}=\frac{1}{\pi}\left[\frac{af(\nu)}{a^2+x^2}\right].
\eeq
and $a\equiv\Gamma/(4\pi\dnd)$. For Lyman $\alpha$, 
$a=6.04\times 10^{-4}/\dvds$, where $\dvds\equiv\dvd/(10^6$~cm~s$^{-1})$.
The line profile for a gas, $\phi_x$, is the same as this in the line wings
($x\gg 1$). We conclude that the line profile
in the damping wings is given by 
\beq
\phi_x\simeq \frac{af(\nu)}{\pi x^2}~~~~~(x\gg 1).
\eeq
The correction to the usual expression for the damping profile
is given by the factor $f(\nu)$, which drops to 0.5 at
$\nu=0.80\nu_0$.
The relation between $\nu$
and $x$ is given by
\beq
\frac{\nu}{\nu_0}=1+\frac{x\dnd}{\nu_0}=1+3.33\times 10^{-5}\dvds x,
\eeq
where the final expression is for Lyman $\alpha$.

        The optical depth at a frequency labeled by $x$
is $\tau_x=\taueff\phi_x$. For Lyman $\alpha$,
\beq
\taueff=1.34\times 10^{-13}N_{\rm eff}/\dvds,
\label{eq:tauefflal}
\eeq
(Neufeld 1990), where $N_{\rm eff}$ 
is the effective column density of
\ion{H}{1} (cf. eq. \ref{eq:harmonic}).
Thus, in the damping wings of Lyman $\alpha$ we have
\beq
\tau_x=2580\left[\frac{N_{\rm eff,20}f(\nu)}{\dvds^2 x^2}\right]
~~~~~~(x\gg 1).
\eeq
The frequency $0.8\nu_0$, where $f(\nu)=0.5$, corresponds to 
$x=6000/\dvds$. In order to have optical depth unity at this
point, the column density must be $N_{\rm eff}\sim 3\times 10^{24}$.

\section{Enhancement of \lal\ Intensity in an Optically Thick Medium}
\label{app:enhance}

Here we derive the increase in \lal\ intensity relative to the
optically thin limit due to the trapping of photons in regions of high
column densities, such as the neutral gas around the protostellar
\ion{H}{2} region. This factor was used in \S\ref{S:lya} for the
calculation of the \lal\ radiation pressure feedback, and some of the
symbols in this appendix are defined there.
We first consider the case of pure scattering, and then include
the effect of destruction by two-photon emission.

\subsection{Case of Pure Scattering}

We follow the treatment of Neufeld (1990),
which extended earlier work by Harrington (1973).
He considered a uniform,
planar slab of thickness $2\bar\tau_L$, with the origin at the center
of the slab.  A planar source of photons located at $\bar\tau_s$
produces 
an incident flux $F_i$ 
in each direction; he normalized to $F_i=1/2$.  We
assume that there is no absorption.  First consider the case in which
the photons are injected at line center ($x_i=0$).  The
frequency-integrated intensity at a point $\bar\tau$ in the slab is
\beq
\frac{J(\bar\tau;\, x_i=0)}{2F_i}
= \left(\frac 32\right)^{1/2}
\left(\frac{4a\bar\tau_L}{\pi}
\right)^{1/3}\left[\calf\left(\frac{\bar\tau-\bar\tau_s'}{2\bar\tau_L}
\right)-\calf\left(\frac{2\bar\tau_L-\bar\tau-
\bar\tau_s'}{2\bar\tau_L} \right)\right], 
\eeq 
where 
\beq
\calf(w)=\frac{\surd 6\Gamma(\frac13)}{12\pi^{7/3}}\sum_{n=1}^\infty
\frac{\cos(n\pi w)}{n^{4/3}}.  
\eeq 
In general,
\beq 
\bar\tau_s'\equiv
\bar\tau_s\left(1-\frac{2}{3\bar\tau_L\phi_i}\right), 
\eeq 
where $\phi_i=\phi(x_i)$ is the line profile at the injection frequency; we
have assumed that the photons are injected 
at line center, so that
$\phi_i=\phi_0$ 
is not small. We shall further assume that
$\bar\tau_s'\simeq\bar\tau_s$, which is valid provided
$\Delta\bar\tau_s\equiv\bar\tau_L-\bar\tau_s\gg 1$---i.e., the source
is not too near the edge of the slab.

        For the case in which the source is at the center of 
the slab $(\bar\tau_s=0)$, the intensity at the center of the slab is
\beq
J_s(x_i=0)=J(0;\, x_i=0)
=\left(2^{4/3}-1\right)
        \left[\frac{\Gamma(\frac13)\zeta(\frac 43)}{2^{2/3}\pi^{8/3}}
        \right](a\bar\tau_L)^{1/3}F_i=0.4362(a\bar\tau_L)^{1/3}F_i,
\label{eq:har}
\eeq
as originally found by Harrington (1973). This has the same scaling
as expected from the heuristic argument 
given in \S\ref{S:lya}:
in the absence of any scattering, the mean intensity would
be $J=F_i/2\pi$ 
(assuming isotropic emission), so the
mean intensity is indeed enhanced by a factor of order
$(a\bar\tau_L)^{1/3}$.

   Next, consider the case in which the source is near the edge of
the slab, 
but, in view of the smallness of the Lyman-$\alpha$ mean free path,
still at a large optical depth from the edge
($\bar \tau\gg \Delta\bar\tau_s\gg 1$).
The maximum intensity
occurs at $\bar\tau_s$, and is proportional to
\beq
\calf(0)-\calf\left(\frac{\Delta\bar\tau_s}{\bar\tau_L}\right)
        =\frac{\surd 6\Gamma(\frac13)}{12\pi^{7/3}}
        \sum_{n=1}^\infty \frac{1-\cos\left(\frac{n\pi \Delta\bar\tau_s}
        {\bar\tau_L}\right)}{n^{4/3}}.
\eeq
Approximating the sum by 
\beq
2\int_0^\infty dn\;n^{-4/3}\sin^2\left(\frac{n\pi\Delta\bar\tau_s}
          {2\bar\tau_L}\right)=
          -\Gamma\left(-\frac 13\right)\cos\left(\frac{\pi}{6}\right)
          \left(\frac{\pi\Delta\bar\tau_s}{\bar\tau_L}\right)^{1/3}
\eeq
(Gradshteyn \& Ryzhik 1965), we find
\beq
J_s(x_i=0)
=0.518(a\Delta\bar\tau_s)^{1/3}F_i
=0.411(a\taueff)^{1/3}F_i,
\label{eq:deltataus}
\eeq
where the effective optical depth is 
$\taueff=2\Delta\bar\tau_s$ (eq. \ref{eq:harmonic})
for the case in which the source
is close to one edge.
Note that this agrees quite well with our {\it ansatz}, equation
(\ref{eq:harmonic}), since when the result for a source near
the edge is expressed in terms of $\taueff$, it is nearly the
same as that for a source near the center (eq. \ref{eq:har}).

      Neufeld's results are valid only in the limit in which
$a\bar\tau_L\ga 10^3$, so that the transfer is completely
determined by the damping wings.  Hummer \& Kunasz (1980)
show that for smaller optical depths, there is an intermediate
regime in which $J/F_i$ is about constant.  They define
a quantity $\rhk\equiv \rho\bar\kappa D/\bar\tau_L$, where 
the density $\rho$ is assumed to be constant and where
$D$ is the mean distance traveled by an escaping photon.
As shown by Ivanov (1970), this is $(4\pi/2\btl F_i)\int J d\bar\tau$,
which in turn is proportional to $J_s/F_i$.
For the case
$1\ga\dvds\ga 0.1$ (corresponding to $10^4$~K$\ga T\ga 10^2$~K), 
the results of
Hummer \& Kunasz (1980) imply that $\rhk$, and therefore
$J_s/F_i$, are within about 0.1 dex of their values at 
$a\bar\tau_L=450$ for $10^3\ga a\bar\tau_L\ga 1$.
We assume that the same behavior obtains in terms of $\taueff$ for
a source near the edge of the slab.
Thus, for $\taueff\ga 1/a=1660\dvds$, we have
\beq
\left.\frac{J_s}{F_i}\right|_{x_i}
        \simeq \frac{0.411(a\taueff)^{1/3}}{{\rm Min}\left[
        1,\;(a\taueff/ 450)^{1/3}\right]}.
\label{eq:max}
\eeq
For $450\ga a\taueff\ga 1$, this gives $J_s/F_i\simeq 3.15$.
The condition
$a\taueff\ga 1$ for the validity of
equation (\ref{eq:max}) corresponds to
to column densities $N_{\rm eff}\ga  10^{16}\dvds^2$~cm\ee.

        This result is valid for \lal\ photons produced by
the \ion{H}{2} region, since such photons are very near line center.
Stellar FUV photons are not restricted to line center, but fortunately  
Neufeld has evaluated the intensity for an arbitrary 
injection frequency. Since his 
results for this case are somewhat
complicated, so we shall evaluate the intensity far from line center
and then smoothly join the result onto the result we have found above.
If the injection frequency is large compared to the diffusion
frequency [$x_i\gg (a\taueff)^{1/3}$], then the photons scatter
approximately coherently. The intensity in this case can be found
either from Neufeld's general results\footnote{The numerical
coefficient in Neufeld's equation (2.30) is too small by a factor
3, as confirmed by the author (private communication).}
or from a simple solution
to the radiative transfer equation.  
We normalize the injection
frequency,
\beq
\hat x_i\equiv \frac{x_i}{(a\taueff)^{1/3}},
\eeq
so that photons injected with $\hat x_i\ll 1$ are in the diffusion
regime and those with $\hat x_i\gg 1$ are in the coherent scattering
regime.
In the latter case, we find
\beq
\left.\frac{J_{s,\,\iso}}{F_i}\right|_{x_i}=\frac{3}{4\pi}\,\phi_i\taueff
        =\left(\frac{3}{4\pi^2}\right)\frac{(a\taueff)^{1/3}f(\nu)}{
        \hat x_i^2}~~~~~(\hat x_i\gg 1),
\label{eq:coherent}
\eeq
where $f(\nu)$ is the Rayleigh-scattering factor defined in equation
(\ref{eq:fnu}).  We have put the subscript ``iso'' on the mean
intensity to indicate that it has been derived under
the assumption that the medium is
optically thick at the frequency $x_i$ so that
the photons are isotropic. Sufficiently far in 
the line wings, equation (\ref{eq:coherent}) shows that
$J_{s,\,\iso}$ goes to zero.  This approximation is
developed further in Appendix \ref{app:radpress} below.
If several lines contribute to 
the opacity at a given frequency, then the right hand side of
equation (\ref{eq:coherent}) should be summed over the lines,
since it is the total opacity that governs the mean intensity.

        We are now in a position to join the result for
the frequency diffusion
(eq. \ref{eq:deltataus}) to that for coherent scattering in the 
far wings of the line (eq. \ref{eq:coherent}).
Taking the harmonic mean of these results, we
obtain an expression that is approximately
valid for all injection frequencies:
\beq
\left.\frac{J_{s,\,\iso}}{F_i}\right|_{x_i}\simeq 
        \frac{0.411 (a\taueff)^{1/3}}{{\rm Min}[1,\;(a\taueff/450)^{1/3}]
        +5.41[\hat{x}_i^2/f(\nu)]}.
\eeq    
Note that the numerical coefficient has been chosen to agree
with the case of a source near the edge of a cloud, which is
the one most relevant to our problem.
The intensity is half that at line center for an injection frequency
$\hat x_i=0.43$ (for $a\taueff>450$ and $f[\nu]\simeq 1$).

\subsection{Effect of Two-Photon Emission}

        In the absence of dust or molecular hydrogen, the
dominant destruction processes for \lal\ photons are 
collisional de-excitation (which we shall ignore)
and two-photon emission following 
a collisional transition from the $2p$ to the $2s$ state.
\lal\ photons can also
be destroyed by photoionization out of the $n=2$ state, but
since another \lal\ photon is created after the ion recombines,
the net destruction by this process vanishes. 

        Consider a 1D slab of gas with a central source of \lal\
photons. Let $\epsilon$ be the probability of two-photon
emission per scattering.  Then in the limit of large optical
depth, the mean intensity at the source is (Harrington 1973)
\beq
J_s(x_i=0)=0.396(a/\epsilon)^{1/3}F_i,
\eeq
which is quite close to the result with no absorption
(eq. \ref{eq:har}) with the replacement $\btl\rightarrow 1/\epsilon$.
Thus, two-photon emission prevents the mean intensity from
increasing once $\epsilon\btl\ga 1$. We join this result onto the
expression for the case in which the source is not at the center and
there is no absorption (eq. \ref{eq:max}) by writing
\beq
J_s(x_i=0)=\frac{0.411(a\taueff)^{1/3}}{{\rm Min}[1,\, (a\taueff)^{1/3}]
        +\Gamma},
\eeq
with
\beq
\Gamma=1.04(\epsilon\taueff)^{1/3}.
\label{eq:gamma1}
\eeq

        To determine the destruction probability $\epsilon$, we
need to know the population of the $2s$ state.
In statistical equilibrium, this is
\beq
\frac{n_{2s}}{n_{2p}}=\frac 13 \left(\frac{1}{1+\necr/n_e}\right),
\eeq
where 
\beq
\necr\equiv \frac{A_{2s1s}}{q_{2s2p}}=1.55\times 10^4~~~~~~{\rm cm^{-3}}
\eeq 
is the critical density for the $2s\rightarrow2p$ transition,
$A_{2s1s}=8.23$~s\e\ is the two-photon emission rate from the
$2s$ state, and $q_{2s2p}=5.31\times 10^{-4}$~cm$^3$~s\e\ is the 
collisional rate coefficient for electron and proton collisions
from the $2s$ state to the $2p$ state (Osterbrock 1989).
Collisional de-excitation from $2s$ to $1s$ is much slower, and
may be neglected in determining this population ratio.
The probability per scattering
of destroying a \lal\ photon by two-photon emission is then
\beq
\epsilon=\frac{n_{2s}A_{2s1s}}{n_{2p}A_{2p1s}}
        =\frac{4.4\times 10^{-9}}{1+\necr/n_e}.
\eeq
Collisional de-excitation to the $1s$ level competes with this
process for $n_e\ga 10^8$~cm\eee, but we shall assume that
$n_e$ is less than this so that we can ignore this process.
Inserting this result into equation (\ref{eq:gamma1}),
we find that the factor that gives the effect of two-photon emission is 
\beq
\Gamma=\frac{0.405(N_{\rm eff,\, 20}/\dvds)^{1/3}}
        {(1+\necr/n_e)^{1/3}}.
\eeq
We see that at high electron densities ($n_e\gg\necr$), two-photon
emission reduces the \lal\ intensity by a factor that depends
only on the column density and the velocity dispersion.  Although
the large column densities of \ion{H}{1} needed to make this process
important occur in regions of neutral hydrogen,
one can show that photoionization out
of the $n=2$ state is generally sufficiently effective that
$n_e\ga\necr$ in the \ion{H}{1} just outside the \ion{H}{2} regions 
of massive primordial stars.  As a result, we have
\beq
\Gamma\simeq 0.405(N_{\rm eff,\, 20}/\dvds)^{1/3}.
\eeq

        This destruction process operates only for photons that
can diffuse to the center of the line, which is where most of the
scatterings take place.  For stellar FUV photons, we assume
that this occurs only for photons within a frequency
range $(a\taueff)^{1/3}\dnd$, or $|\hat x_i|<1$,
so that $\Gamma$ depends
on $\hat x_i$:
\beq
\Gamma(\hat x_i)=\left\{\begin{array}{r@{\qquad \qquad}l}
        \Gamma & |\hat x_i|<1, \\
        0       &    |\hat x_i|>1.
        \end{array}\right.
\eeq    

\subsection{The Blackbody Constraint\label{app:enhancebb}}

        We have one last constraint to impose: the intensity
we calculate must be less than 
the appropriate blackbody intensity $B_\nu$. 
The \lal\ photons produced by the \ion{H}{2} region have a complicated
line profile that is concentrated in a frequency range
$\sim 2(a\taueff)^{1/3}\dnd$.  For these photons, we require that the
intensity in this frequency range be less than that of a 
blackbody at the temperature of the \ion{H}{2} region,
\beq
J_{\alpha,\,\rm HII}={\rm Min}\left\{2(a\taueff)^{1/3}\dnd
        B_{\nu_\alpha}(T_{\rm HII}),
        \frac{0.411 (a\taueff)^{1/3}F_{\alpha,\,\rm HII}}
        {{\rm Min}[1,\;(a\taueff/450)^{1/3}]+\Gamma}\right\}.
\label{eq:jhii}
\eeq

        For stellar photons,
the intensity is limited by the blackbody intensity
at the stellar surface, $B_\nu(T_*)
=F_{\nu *}/\pi$:
\beq
        J_{\alpha*,\,\iso}=\int d\nu_i\;
        {\rm Min}\left\{B_{\nu_i}(T_*),\;
        \frac{0.411 (a\taueff)^{1/3}F_i(\nu_i)}
        {{\rm Min}[1,\;(a\taueff/450)^{1/3}]
        +5.41[\hat{x}_i^2/f(\nu)]+\Gamma(\hat x_i)}\right\}.
\label{eq:j*}
\eeq
The subscript ``iso'' indicates that $J_{\alpha *,\,\iso}$
is that part of the stellar radiation that has been isotropized
by scattering (see Appendix \ref{app:radpress} below).
The factor in the denominator ensures that although the
integral is taken over the entire spectrum of the star, it
is only that part near the resonance line that contributes.

\section{Estimate of the Radiation Pressure}
\label{app:radpress}

        In opaque media, the force exerted by radiation can 
be treated as an isotropic pressure, just like gas pressure.
If the medium is not opaque, however, the radiation pressure
tensor is not diagonal, and radiation pressure does not behave
like gas pressure.  For example, in the optically thin limit,
the radiation pressure declines as $r^{-2}$, yet the associated
force per unit volume is negligible.  Here we introduce
an approximation to the radiation pressure tensor that
separates out the isotropic part and show that the radiative force
is the gradient of this pressure.

        The radiative force per unit volume is
\beq
\vecfrad=-\vecnabla\cdot\rvecprad=\frac{1}{c}\, \rho\kappa \vecF,
\eeq
where $\rvecprad$ is the radiation pressure tensor (e.g., Shu 1991);
for a frequency-dependent opacity,
$\kappa\vecF=\int\kappa_\nu\vecF_\nu$.
At large optical depths, the radiation field is nearly
isotropic and 
the radiation pressure tensor
becomes diagonal, $\rvecprad\rightarrow
\pradi{\bf I}$.
At small optical depths, the radiation
is beamed and $\rvecprad\rightarrow (1/c) \vecF\vecFhat$, where
$\vecFhat$ is a unit vector.  (Note that in one dimension,
the radiation is not purely beamed at small optical depth, 
so there is a numerical
coefficient $\leq 1$ in front of $\vecF$ in this expression.) 
We then have for all optical depths
\beq
\rvecprad\simeq \pradi {\bf I}+\frac{1}{c}\,\vecF\vecFhat,
\eeq
and the radiative force is 
\beq
\vecfrad= -\vecnabla \cdot\rvecprad \simeq -\vecnabla \pradi -
\frac{1}{c}
\vecFhat\vecnabla\cdot \vecF.
\eeq
Here we have omitted the term $\vecF\cdot\vecnabla\vecFhat$, which
vanishes in the optically thin limit since then the flux does not change
direction along a ray, and which is negligible in the optically thick
limit even if the flux does change direction.  In the case of
pure scattering, or in a stellar atmosphere, we have
$\vecnabla\cdot\vecF=0$, so the radiative force is then
\beq
\vecfrad\simeq -\vecnabla\pradi=\frac{1}{c}\,\rho\kappa \vecF
        ~~~~~~~({\rm for~}\vecnabla\cdot\vecF=0).
\label{eq:prad}
\eeq
This is the case relevant to our problem.
On the other hand, in the case of strong absorption, so that
$F$ decreases in a distance small compared to the radius, 
one cannot neglect the term $\vecFhat\vecnabla\cdot \vecF$
in evaluating the force.

        If $\vecF$ is known, then $\pradi$ can be determined from
equation (\ref{eq:prad}) once a boundary condition is specified.
We assume that there is a surface to the gas distribution.
It follows that $\vecfrad\propto \kappa$ vanishes
outside the surface, so that $\pradi$ must
be constant there. Since the radiation pressure vanishes at
infinity, it follows that the constant must be zero.
We conclude that $\pradi=0$ at the surface of the gas 
distribution.

        At large optical depths, the energy density
of the radiation is $\urad=4\pi J/c=3\prad$.  We therefore
define $\uradi\equiv 4\pi J_\iso/c \equiv 3\pradi$.
Since $\urad\rightarrow F/c$ at small optical depths,
we have $\urad\simeq \uradi+F/c$.

        Under the assumptions
of spherical symmetry and no net absorption
($\vecnabla\cdot\vecF=0$), our formulation is equivalent
to the closure approximation recommended by Shu (1991, pp. 43-44).
He shows that the equation for the $rr$ component of the radiation
pressure tensor is
\beq
\frac{\partial \pradrr}{\partial r}+\frac{1}{r}(3\pradrr-\urad)
        =-\frac{1}{c}\,\rho\kappa F.
\label{eq:shu}
\eeq
Shu points out that $3\pradrr-\urad\simeq 2F/c$, since at
high optical depths $F$ is negligible and
$\prad=\urad/3$, whereas at low optical depths $\pradrr=\urad
=F/c$.  With this approximation, the left-hand side of the
equation becomes $\partial \pradrr/\partial r+2F/cr$.
Our formulation gives $\pradrr=\pradi+F/c$.  Since we have assumed
$\vecnabla\cdot \vecF=0$, we have $\partial F/\partial r=-2F/r$
in the spherically symmetric case, 
so that equation (\ref{eq:shu}) reduces to equation (\ref{eq:prad}).

Recall that the \lal\ radiation pressure, $\pradi$ 
in equation (\ref{eq:pfinal}), was derived
for a slab geometry.
What is the appropriate value of the column density
$N_{\rm eff}$ to use in this expression in a more realistic
geometry? Here we shall determine the relation between
a slab geometry and a spherical one; in the next Appendix,
we generalize this to non-spherical geometries.
However, it must be borne in mind that
the actual geometry of the infalling gas is far more complicated
than can be represented by a simple analytic model.

What is the relation between the optical depth in one dimension
and that in three dimensions that results in the same radiation pressure?
Let $F\propto r^{-k_F}$, with $k_F=0$ for slab geometry and
$k_F=2$ for spherical geometry. We assume
that the density in the spherical case can
be described by a power law, $\rho\propto r^{-\krho}$.  
For a constant opacity per unit mass,
we therefore
have
\beq
\frac{\partial\pradi}{\partial r}=-\frac{\rho_0\kappa F_0}{c}\left(
        \frac{r_0}{r}\right)^{k_F+\krho},
\label{eq:sphere}
\eeq
where $r_0$ is a fiducial radius and $\rho_0\equiv\rho(r_0)$, etc.
The optical depth from $r$ to infinity in the spherical case is
\beq
\tau_{\rm 3D}=\frac{\rho\kappa r}{\krho-1}
\eeq
for $\krho>1$,
so that the solution of equation (\ref{eq:sphere}) is
\beq
\pradi=\left(\frac{\krho-1}{\krho+k_F-1}\right)\frac{F\tau}{c}.
\eeq
In order for the radiation pressure in the slab to be
the same as that in a sphere for the same value of the flux,
we require
\beq
\tau_{\rm 1D}=\left(\frac{\krho-1}{\krho+1}\right)\tau_{\rm 3D}
=\frac{\rho\kappa r}{\krho+1}.
\eeq
Since $\tau_{\rm 1D}=\muh\kappa N_{\rm eff}$, 
we conclude that 
\beq
N_{\rm eff} = \frac{ n r} {k_\rho +1} 
\eeq 
For a free-fall density variation, valid for $r\gtrsim r_d$,
we have $\krho=3/2$ so that $N_{\rm eff}= (2/5) nr$; inside $r_d$,
$k_\rho=1/2$ is a more accurate description, so that $N_{\rm eff}= (2/3) nr$.

\section{Anisotropic Optical Depth and Super-Critical Accretion}
\label{app:aniso}

        In order to estimate how Lyman-$\alpha$ photons escape
from the \ion{H}{2} region around the protostar, we consider the
following idealized problem: We assume that the radiation
fills a  cavity bounded by a thin, opaque shell
of variable optical depth, $\tau({\bf r})$.
In this case, the flux at the surface of the shell
is approximately normal to the surface, 
$\vecFhat\simeq \vecnhat,$
and the radiation energy density is about
constant in the interior of the shell. This model will
be approximately
valid for \lal\ radiation
once the \ion{H}{2} region separates from the star,
since the \ion{H}{2} region provides a cavity in which the 
optical depth due to resonance line scattering is relatively
small, so that the
radiation becomes
approximately uniform there.
Equation (\ref{eq:prad}) then gives
\beq
\vecF({\bf r})\simeq -c\frac{d\pradi}{d\tau({\bf
        r})}\;\vecnhat\simeq\frac{c\pradi}{\tau({\bf r})}\;\vecnhat.
\label{eq:fr}
\eeq
Integration over the surface of the shell
gives the luminosity:
\beq
L=\int {\bf F\cdot \hat n}\,dA\simeq
         c\pradi\int\frac{dA}{\tau({\bf r})}
        = \frac{c\pradi A}{\taueff},
\label{eq:L2}
\eeq
where $A$ is the total area of the shell and
$\taueff$ is the the harmonic mean
optical depth:
\beq
\frac{1}{\taueff}\equiv\frac{1}{A}\int\frac{dA}{\tau({\bf
        r})}.
\eeq
For a spherical shell, this simplifies to
\beq
\frac{1}{\taueff}=\frac{1}{4\pi}\int\frac{d\Omega}{\tau({\bf
        \hat r})}.
\eeq

        We can generalize this treatment to allow for the possibility
that the optical depth is small in some directions. Consider
the extreme case in which $\tau=0$ over a small area $\delta
A$---i.e., there is a small hole in the shell.
The flux emerging from this area is $\pi$ times the
specific intensity, which is the same as the mean intensity
$J$ in the cavity. We therefore find
\beq
F({\bf r})=\pi J=\frac{c\urad}{4}=\frac{3c\prad}{4}\simeq\frac{3c\pradi}{4}
        ~~~~~(\tau=0),
\eeq
where the last step follows since we have assumed that the
average optical depth is large enough that $\prad\gg F/c$
so that $\prad\simeq \pradi$ 
(note that $\pradi$ drops near the hole, but that does not
affect the average value of $\pradi$ since the hole is small) .
Combining this result with equation (\ref{eq:fr}), we write
\beq
F({\bf r})\simeq\frac{c\pradi}{\tau({\bf r})+\frac 43}
\eeq
as an expression that is approximately valid for all $\tau$.
With
\beq
\frac{1}{\taueff}\equiv\frac{1}{A}\int\frac{dA}{\tau({\bf
        r})+\frac 43},
\eeq
equation (\ref{eq:L2}) is valid even if the optical depth
is small in some directions. As an example, assume that
$\tau=0$ over an area $\delta A$ and
$\tau=\tau_0 \gg 1$
elsewhere. Then we have
\beq
\taueff=\frac{\tau_0}{\displaystyle \left(1-\frac{\delta A}{A}
        \right)+\frac{\delta A}{A}\left(\frac{\tau_0}{4/3}\right)}.
\eeq
We require $\taueff\gg 1$ in order for our treatment to be valid,
and this will be true if both $\tau_0\gg 1$ and $\delta A/A\ll 1$.

        Equations (\ref{eq:fr}) and (\ref{eq:L2})
imply that the flux at any point on the shell is
then 
\beq
F({\bf\hat r})=\frac{\taueff}{\tau({\bf r})}\left(\frac{L}{A}\right).
\eeq
In our problem, $\tau<\taueff$ near the poles, 
so the flux there is enhanced over 
$L/A$ since radiation originally directed at
regions of high optical depth tends to escape in
regions of low optical depth. This is a quantitative
expression for the flashlight effect (Yorke \&
Bodenheimer 1999).

        When radiation pressure is acting against gravity,
it is convenient to define the
critical flux as the flux that just
counterbalances gravity,
\beq
F_\crit=\frac{Gm_*c}{r^2\kappa},
\eeq
where we have assumed spherical symmetry and where
$\kappa =\sigma/\mu$ is the opacity per unit mass
and $\mu$ is the mean mass per particle.
The critical luminosity is then
\beq
L_\crit\equiv 4\pi r^2 F_\crit=\frac{4\pi Gm_*\mu c}{\sigma}.
\eeq
If the opacity is due to electron scattering, the critical
luminosity is the Eddington limit.
We conclude that supercritical accretion---i.e., accretion when
$L>L_\crit$---can occur
in directions with $\tau>\taueff$
since it is possible for $F$ to be less than $F_\crit$ in
those directions:
\beq
\frac{F({\bf r})}{F_\crit}=
\frac{\taueff}{\tau({\bf r})}\left(\frac{L}{L_\crit}\right).
\eeq
For example, an accretion disk can produce supercritical
accretion since the optical depth in the plane of the disk
is much larger than that in other directions.

        This argument works well in our problem because
the \lal\ opacity  in the central regions is small due to 
photoionization, thereby rendering the radiation approximately
uniform there.
It is more difficult to create a uniform radiation
field in the case
of electron scattering in an ionized gas, 
since the opacity per unit mass is constant
and it is difficult to create a thin, opaque shell 
around a star. Nonetheless,
the effective optical depth in this case is likely to
be of order the harmonic mean optical depth, just as we
have found for our idealized problem.

\section{VERTICAL STRUCTURE OF AN ACCRETION DISK SUPPORTED BY
GAS PRESSURE, WITH CONSTANT OPACITY}
{\label{A:vdisk}}

Here we determine the height of an accretion disk
supported by gas pressure
under the assumption that
the opacity per unit mass, $\kappa$, is constant.
For a disk supported by gas pressure, the equations describing
the radiation field in the disk 
(see \S \ref{S:shadow}) can be written as:
\beq
\frac{dT^4}{d\Sigma}=\frac{3\kappa F}{4\sigma}
\label{eq:t4}
\eeq
(radiative diffusion), where 
\beq
\Sigma\equiv\int_z^{z_s}\rho dz
\label{eq:Sigma}
\eeq
is the surface density above a height $z$; and 
\beq
\frac{dF}{dz}=\frac{F_0P}{\int_0^{z_s}Pdz}
\label{eq:dfdzapp}
\eeq
(flux generation).
Since gas pressure dominates, we have $P=\rho kT/\mu$,
so that this becomes
\beq
\frac{dF}{d\Sigma}=-\frac{F_0 T}{\Sigma_c \avg{T}},
\label{eq:dfds}
\eeq
where $\Sigma_c=\int_0^{z_s}\rho dz$ is half the total surface density
of the disk and $\avg{T}$ is the mass-weighted average temperature in 
the disk.

    To obtain an approximate solution for these two equations,
we set
\beq
T\simeq
T_c\ssc^{1/4}\left[\frac{1+\epsilon(\Sigma/\Sigma_c)^\ell}{1+\epsilon}
        \right],
\label{eq:tansatz}
\eeq
where $T_c$ is the central temperature.
The parameters $\epsilon$ and $\ell$ are to be determined; in particular,
$\epsilon$ is assumed to be small, so that
\beq
T^4\simeq T_c^4 \ssc \left[\frac{1+4\epsilon(\Sigma/\Sigma_c)^\ell}
          {1+4\epsilon}\right].
\eeq
Inserting this into equation (\ref{eq:t4}), we find
\beq
F=\frac{4\sigma T_c^4}{3\tau_c}\left[\frac{1+4(\ell+1)
        \epsilon(\Sigma/\Sigma_c)^\ell}{1+4\epsilon}\right],
\eeq
where $\tau_c\equiv \kapr\Sigma_c$ is the optical depth from
the midplane to the surface.
Since $F=0$ at the midplane, where $\Sigma=\Sigma_c$, we find
\beq
\epsilon=-\frac{1}{4(\ell+1)}.
\eeq
At the surface ($\Sigma=0$), we have $F=F_0$, so that
\beq
F_0=\frac{1}{(1+4\epsilon)}\;\frac{4\sigma T_c^4}{3\tau_c},
\label{eq:fo}
\eeq
and
\beq
F=F_0\left[1-\ssc^\ell\right].
\label{eq:fssc}
\eeq
Inserting this into equation (\ref{eq:dfds}) 
and keeping only the leading term in equation (\ref{eq:tansatz})
implies $\ell=5/4$, so that $\epsilon=-1/9$. Since $F_0=\sigma\teff^4$,
equation (\ref{eq:fo}) implies
\beq
T_c=\left(\frac {5}{12}\; \tau_c\right)^{1/4}\teff.
\label{eq:tc}
\eeq
This approach gives a mass-weighted mean temperature 
$\avg{T}=\frac 45 T_c$ from equations (\ref{eq:dfds}) and (\ref
{eq:fssc});
on the other hand, direct integration of equation (\ref{eq:tansatz})
gives $\avg{T}=\frac{17}{20}T_c$. The 6\% difference between
these estimates for $\avg{T}$ is a measure of the accuracy
of our approximations.

        The equation of hydrostatic equilibrium is
\beq
\frac{dP}{d\Sigma}=\frac{ g_0 z}{\varpi}.
\label{eq:hseapp}
\eeq
Approximating $T\simeq T_c(\Sigma/\Sigma_c)^{1/4}$, which is
typically accurate to better than 10\%, we then find
\beq
P=\rho \cgc^2\left(\frac{\Sigma}{\Sigma_c}\right)^{1/4}
       =\frac{g_0}{\varpi}\int_0^\Sigma zd\Sigma'.
\label{eq:P}
\eeq
Define the characteristic scale height as
\beq
h_{gc}\equiv\frac{\cgc^2}{g_0}=\left(\frac{5\tau_c}{12}\right)^{1/4}
        \frac{k\teff}{\mu g_0},
\eeq
where the second step follows from equation (\ref{eq:tc}).
In terms of the Keplerian velocity
$v_K=(g_0\varpi)^{1/2}$, the scale height is $h_{gc}/\varpi=(c_{gc}/v_K)^2$.
Since $\rho=-d\Sigma/dz$, equation (\ref{eq:P}) yields the following
equation for $\Sigma$:
\beq
\left(\frac{\Sigma}{\Sigma_c}\right)^{1/4}\frac{d\Sigma}{dz}=
        -\frac{1}{\varpi h_{gc}}\int_0^\Sigma z d\Sigma'.
\label{eq:dsigmadz}
\eeq
To obtain an approximate solution to this equation, we
adopt the following {\it ansatz} for $\Sigma$:
\beq
\ssc^{1/4}\simeq\left(1-\frac{z}{z_s}\right)\left(1+\frac{z}{2z_s}\right).
\eeq
An approximate evaluation of the integral $\int zd\Sigma'$ gives
\beq
\int zd\Sigma'\simeq z_s\Sigma\left(\frac 13 +\frac{2z}{3z_s}\right).
\eeq
Integration of equation (\ref{eq:dsigmadz}) then gives
\beq
\ssc^{1/4}\simeq \frac{z_s^2}{6\varpi h_{gc}}\left(1-\frac{z}{z_s}\right)
                 \left(1+\frac{z}{2z_s}\right),
\label{eq:soln}
\eeq
which is consistent with the {\it ansatz} provided that
the height of the disk is
\beq
z_{sg}=\left(6\varpi h_{gc}\right)^{1/2},
\label{eq:zsg}
\eeq
where the subscript $g$ indicates that the height is evaluated
for the case in which gas pressure dominates.
Shakura \& Sunyaev (1973) show that the height of the
disk $\sim (c_c/v_K)\varpi$, where $c_c$ is the central isothermal
sound speed.
Equation (\ref{eq:zsg}) implies that in a gas-pressure
dominated disk,
\beq
\frac{z_{sg}}{\varpi}=\frac{\surd 6 c_{gc}}{v_K}.
\label{eq:zsgvk}
\eeq
In terms of the sound speed at the photosphere, $c_{g,\,\rm eff}=
(k\teff/\mu)^{1/2}$, this is
\beq
\frac{z_{sg}}{\varpi}=(540\tau_c)^{1/8}\;\frac{c_{g,\,\rm eff}}{v_K}.
\eeq
Numerical integration of the structure equations shows that the
actual height of a gas-pressure dominated disk
ranges from $1.04 z_{sg}$ for $\tau_c=10^4$ to
$1.10 z_{sg}$ for $\tau_c= 10^9$, so the approximations made in
our analytic estimate are reasonably good.

        Paczynski \& Bisnovati-Kogan (1981) obtained equation 
(\ref{eq:zsgvk})
through a different argument: They assumed that the disk is
polytropic, with $P\propto \rho^{1+1/n}$, and found that
\beq
\frac{z_{sg}}{r}=\frac{[2(n+1)]^{1/2} c_{gc}}{v_K}.
\eeq
They argued that $n$ is likely to be between 1.5 and 3, so
that $2(n+1)$ is between 5 and 8; they chose 6 as a typical
value. We emphasize that our derivation depends only
on the assumption that the opacity is constant, and is
not based on the assumption that the gas is polytropic.

        To complete the determination of the height of a gas-pressure supported disk,
we must estimate the optical depth through half the disk, $\tau_c=
\kapr\Sigma_c$. Observe that
\beq
\int_0^{z_s} P dz=\int_0^{z_s}\left(\frac{kT}{\mu}\right) d\Sigma
             \simeq \left(\frac{k\avg{T}}{\mu_c}\right)\Sigma_c,
\eeq
where $\mu_c$ is the central value of the mean molecular weight.
According to the
discussion below equation (\ref{eq:tc}), the average temperature
is $\avg{T}\simeq \frac 45 T_c$, and we adopt this value here. The fact
that $\avg{T}$ is so close to $T_c$ justifies setting the
mean molecular weight equal to $\mu_c$, as we 
have done.
From equation (\ref{eq:angmom}), we then find
\beq
\tau_c=\frac{\Omega\kapr\mds f}{6\pi\alpha (\frac 45 kT_c/\mu_c)}.
\eeq
Equation (\ref{eq:tc}) then implies
\beq
\frac{T_c}{\teff}=\left[\frac{25\Omega\kapr\mds f}{288\pi\alpha(k\teff/
        \mu_c)}\right]^{1/5}.
\eeq
Using this result in equation (\ref{eq:zsgvk}) for the height of
a gas-pressure supported disk, we find
\beq
z_{sg}=1.21\times 10^{10}\left(\frac{\phi_I\kapr}{\alpha_{-2}\kappa_T}
                \right)^{1/10}\left(\frac{\varpi}{R_\odot}\right)^{21/20}
                \frac{(\mdst f)^{1/5}}{\mst^{7/20}}~~~~~{\rm cm},
\eeq
where we have normalized $\alpha$ to a typical value of 0.01.

\end{document}